\documentclass[fleqn,usenatbib]{mnras}

\usepackage{newtxtext,newtxmath}

\usepackage[T1]{fontenc}
\usepackage{fontawesome}

\DeclareRobustCommand{\VAN}[3]{#2}
\let\VANthebibliography\thebibliography
\def\thebibliography{\DeclareRobustCommand{\VAN}[3]{##3}\VANthebibliography}

\DeclareRobustCommand{\DA}[3]{#2}
\let\DAthebibliography\thebibliography
\def\thebibliography{\DeclareRobustCommand{\DA}[3]{##3}\DAthebibliography}

\usepackage{graphicx}	
\usepackage{amsmath}	

\newcommand{\pyaneti}{\href{https://github.com/oscaribv/pyaneti}{\texttt{pyaneti}\,\faGithub}}
\newcommand{\logr}{$\log R'_{\rm HK}$}
\newcommand{\sshk}{$S_{\rm HK}$}
\newcommand{\lbe}{$\lambda_{\rm e}$}
\newcommand{\lbp}{$\lambda_{\rm p}$}
\newcommand{\pgp}{$P_{\rm GP}$}
\newcommand{\gcm}{${\rm g\,cm^{-3}}$}
\newcommand{\ms}{${\rm m\,s^{-1}}$}
\newcommand{\kms}{${\rm km\,s^{-1}}$}

\newcommand{\citlalicue}{\texttt{citlalicue}}

\newcommand{\vsini}{$v \sin i$}
\newcommand{\logg}{$\log g$}
\newcommand{\ktwo}{\emph{K2}}
\newcommand{\halpha}{$\mathrm{H}_{\alpha}$}
\newcommand{\sodium}{$\ion{Na}{}$}

\newcommand{\tess}{\emph{TESS}}
\newcommand{\cheops}{\emph{CHEOPS}}

\newcommand{\target}{K2-233}
\newcommand{\targetb}{K2-233\,b}
\newcommand{\targetc}{K2-233\,c}
\newcommand{\targetd}{K2-233\,d}
\newcommand{\targetbcd}{K2-233\,b, c, and d}

\newcommand{\Tzerob}[1][days]{ $ 7991.6897 \pm 0.0013 $~#1 } 
\newcommand{\Pb}[1][days]{ $ 2.467583_{-0.000067}^{+0.000062} $~#1 }

\newcommand{\bb}[1][ ]{ $ 0.20_{-0.12}^{+0.11} $~#1 } 
 
\newcommand{\rrb}[1][ ]{ $ 0.01698_{-0.00038}^{+0.00041} $~#1 } 
\newcommand{\kb}[1][${\rm m\,s^{-1}}$]{ $ 1.31_{-0.74}^{+0.81} $~#1 } 
\newcommand{\mpb}[1][$M_{\oplus}$]{ $ 2.4_{-1.3}^{+1.5} $~#1 } 
\newcommand{\rpb}[1][$R_{\oplus}$]{ $ 1.315 \pm 0.036 $~#1 }

\newcommand{\ib}[1][deg]{ $ 88.86_{-0.66}^{+0.72} $~#1 } 
\newcommand{\arb}[1][ ]{ $ 9.97 \pm 0.14 $~#1 } 
\newcommand{\ab}[1][AU]{ $ 0.03293 \pm 0.00066 $~#1 }

\newcommand{\insolationb}[1][${\rm F_{\oplus}}$]{ $ 269 \pm 13 $~#1 } 
\newcommand{\tsmb}[1][ ]{ $ 6.6_{-2.6}^{+8.7} $~#1 } 
\newcommand{\denstrb}[1][${\rm g\,cm^{-3}}$]{ $ 3.08 \pm 0.13 $~#1 } 
 
\newcommand{\Teqb}[1][K]{ $ 1127 \pm 14 $~#1 } 
\newcommand{\ttotb}[1][hours]{ $ 1.882_{-0.047}^{+0.037} $~#1 }

\newcommand{\denpb}[1][${\rm g\,cm^{-3}}$]{ $ 5.7_{-3.2}^{+3.7} $~#1 } 
\newcommand{\grapb}[1][${\rm cm\,s^{-2}}$]{ $ 1334_{-757}^{+832} $~#1 } 
\newcommand{\grapparsb}[1][${\rm cm\,s^{-2}}$]{ $ 1343_{-762}^{+845} $~#1 } 
 
\newcommand{\Tzeroc}[1][days]{ $ 7586.862_{-0.028}^{+0.027} $~#1 } 
\newcommand{\Pc}[1][days]{ $ 7.06024_{-0.00043}^{+0.00044} $~#1 }

\newcommand{\bc}[1][ ]{ $ 0.15_{-0.10}^{+0.12} $~#1 } 
\newcommand{\rrc}[1][ ]{ $ 0.01642 \pm 0.00044 $~#1 } 
\newcommand{\kc}[1][${\rm m\,s^{-1}}$]{ $ 1.81_{-0.67}^{+0.71} $~#1 } 
\newcommand{\mpc}[1][$M_{\oplus}$]{ $ 4.6_{-1.7}^{+1.8} $~#1 } 
\newcommand{\rpc}[1][$R_{\oplus}$]{ $ 1.272 \pm 0.038 $~#1 }

\newcommand{\ic}[1][deg]{ $ 89.57_{-0.35}^{+0.29} $~#1 } 
\newcommand{\arc}[1][ ]{ $ 20.10 \pm 0.28 $~#1 } 
\newcommand{\ac}[1][AU]{ $ 0.0664 \pm 0.0013 $~#1 }

\newcommand{\insolationc}[1][${\rm F_{\oplus}}$]{ $ 66.2 \pm 3.3 $~#1 } 
\newcommand{\tsmc}[1][ ]{ $ 2.1_{-0.7}^{+1.3} $~#1 }

\newcommand{\Teqc}[1][K]{ $ 794 \pm 10 $~#1 } 
\newcommand{\ttotc}[1][hours]{ $ 2.687_{-0.064}^{+0.051} $~#1 }

\newcommand{\denpc}[1][${\rm g\,cm^{-3}}$]{ $ 12.4_{-4.6}^{+5.1} $~#1 } 
\newcommand{\grapc}[1][${\rm cm\,s^{-2}}$]{ $ 2792_{-1025}^{+1148} $~#1 } 
\newcommand{\grapparsc}[1][${\rm cm\,s^{-2}}$]{ $ 2824_{-1044}^{+1143} $~#1 } 
 
\newcommand{\Tzerod}[1][days]{ $ 8005.5802_{-0.0014}^{+0.0016} $~#1 } 
\newcommand{\Pd}[1][days]{ $ 24.3681_{-0.0014}^{+0.0013} $~#1 } 
\newcommand{\ed}[1][ ]{ $ 0.18 \pm 0.09 $~#1 } 
\newcommand{\wdd}[1][deg]{ $ 168_{-25}^{+20} $~#1 } 
\newcommand{\bd}[1][ ]{ $ 0.28_{-0.18}^{+0.16} $~#1 } 
\newcommand{\rrd}[1][ ]{ $ 0.03051_{-0.00065}^{+0.00069} $~#1 } 
\newcommand{\kd}[1][${\rm m\,s^{-1}}$]{ $ 2.72_{-0.70}^{+0.66} $~#1 } 
\newcommand{\mpd}[1][$M_{\oplus}$]{ $ 10.3_{-2.6}^{+2.4} $~#1 } 
\newcommand{\rpd}[1][$R_{\oplus}$]{ $ 2.363 \pm 0.062 $~#1 } 
\newcommand{\Tperid}[1][days]{ $ 8009.27_{-0.86}^{+1.43} $~#1 } 
 
\newcommand{\id}[1][deg]{ $ 89.62_{-0.18}^{+0.24} $~#1 } 
\newcommand{\ard}[1][ ]{ $ 45.91 \pm 0.65 $~#1 } 
\newcommand{\ad}[1][AU]{ $ 0.1516 \pm 0.0030 $~#1 }

\newcommand{\insolationd}[1][${\rm F_{\oplus}}$]{ $ 12.7 \pm 0.6 $~#1 } 
\newcommand{\tsmd}[1][ ]{ $ 27.1_{-5.4}^{+9.0} $~#1 }

\newcommand{\Teqd}[1][K]{ $ 525 \pm 6 $~#1 } 
\newcommand{\ttotd}[1][hours]{ $ 3.791_{-0.085}^{+0.102} $~#1 }

\newcommand{\denpd}[1][${\rm g\,cm^{-3}}$]{ $ 4.3 \pm 1.1 $~#1 } 
\newcommand{\grapd}[1][${\rm cm\,s^{-2}}$]{ $ 1794_{-443}^{+437} $~#1 } 
\newcommand{\grapparsd}[1][${\rm cm\,s^{-2}}$]{ $ 1811_{-451}^{+443} $~#1 } 
 
\newcommand{\qone}[1][]{ $ 0.35_{-0.20}^{+0.33} $~#1 } 
\newcommand{\qtwo}[1][]{ $ 0.32_{-0.19}^{+0.34} $~#1 }

\newcommand{\RV}[1][${\rm km\,s^{-1}}$]{ $ -9.6484 \pm 0.0026 $~#1 } 
\newcommand{\FWHM}[1][${\rm km\,s^{-1}}$]{ $ 7.608 \pm 0.018 $~#1 } 
\newcommand{\BIS}[1][${\rm km\,s^{-1}}$]{ $ 0.0386 \pm 0.0011 $~#1 } 
\newcommand{\jRV}[1][${\rm m\,s^{-1}}$]{ $ 1.10_{-0.72}^{+0.71} $~#1 } 
\newcommand{\jFWHM}[1][${\rm m\,s^{-1}}$]{ $ 9.4_{-2.1}^{+2.0} $~#1 } 
\newcommand{\jBIS}[1][${\rm m\,s^{-1}}$]{ $ 6.66_{-0.57}^{+0.61} $~#1 } 
\newcommand{\jtr}[1][]{ $ 81 \pm 1.5 $~#1 } 
\newcommand{\jAzero}[1][]{ $ 6.5_{-1.5}^{+2.0} $~#1 } 
\newcommand{\jAone}[1][]{ $ 28.2_{-5.0}^{+7.0} $~#1 } 
\newcommand{\jAtwo}[1][]{ $ 44.8_{-8.1}^{+11.3} $~#1 } 
 
\newcommand{\jAfour}[1][]{ $ -1.9_{-1.5}^{+1.4} $~#1 } 
\newcommand{\jAfive}[1][]{ $ -29.1_{-7.5}^{+5.2} $~#1 } 
\newcommand{\jlambdae}[1][]{ $ 17.95_{-1.98}^{+2.08} $~#1 } 
\newcommand{\jlambdap}[1][]{ $ 0.490_{-0.042}^{+0.046} $~#1 } 
\newcommand{\jPGP}[1][]{ $ 9.746_{-0.070}^{+0.065} $~#1 } 

\title[Revisiting K2-233 with multi-GPs]{Revisiting K2-233 spectroscopic time-series with multidimensional Gaussian Processes}

\author[Barrag\'an et al.]{
    Oscar~Barragán$^{1}$\thanks{\href{mailto:oscar.barrragan@physics.ox.ac.uk}{oscar.barrragan@physics.ox.ac.uk} \href{https://twitter.com/oscaribv}{\faTwitter\texttt{@oscaribv}}},
Edward~Gillen$^{2}$,
Suzanne~Aigrain$^{1}$,
Annabella~Meech$^{1}$,
Baptiste~Klein$^{1}$, \and
Louise~Dyregaard~Nielsen$^{1,3}$, 
Haochuan~Yu$^{1}$,
Niamh K. O'Sullivan$^{1}$,
Belinda A. Nicholson$^{1,4}$, \and
and Jorge~Lillo-Box$^{5}$
\\ 
$^{1}$ Sub-department of Astrophysics, Department of Physics, University of Oxford, Oxford, OX1 3RH, UK  \label{oxford} \\
$^{2}$ Astronomy Unit, Queen Mary University of London, Mile End Road, London E14NS, UK \\
$^{3}$ European Southern Observatory | ESO, Karl-Schwarzschild-Str. 2, 85748 Garching bei M{\"u}nchen, Germany\\
$^{4}$ University of Southern Queensland, Centre for Astrophysics, West Street, Toowoomba, Australia, 4350\\
$^{5}$ Centro de Astrobiolog\'ia (CAB, CSIC-INTA), Depto. de Astrof\'isica, ESAC campus, 28692 Villanueva de la Ca\~nada (Madrid), Spain
}

\date{Accepted XXX. Received YYY; in original form ZZZ}

\pubyear{2023}

\begin{document}
\label{firstpage}
\pagerange{\pageref{firstpage}--\pageref{lastpage}}
\maketitle

\begin{abstract}
Detecting planetary signatures in radial velocity time-series of young stars is challenging due to their inherently strong stellar activity. 
However, it is possible to learn information about the properties of the stellar signal by using activity indicators measured from the same stellar spectra used to extract radial velocities.
In this manuscript, we present a reanalysis of spectroscopic HARPS data of the young star K2-233, which hosts three transiting planets. We perform a multidimensional Gaussian Process regression on the radial velocity and the activity indicators to characterise the planetary Doppler signals. We demonstrate, for the first time on a real dataset, that the use of a multidimensional Gaussian Process can boost the precision with which we measure the planetary signals compared to a one-dimensional Gaussian Process applied to the radial velocities alone.
We measure the semi-amplitudes of K2-233\,b, c, and d as \kb, \kc\, and \kd, which translates into planetary masses of \mpb[], \mpc[], and \mpd, respectively.
These new mass measurements make K2-233\,d a valuable target for transmission spectroscopy observations with \emph{JWST}.
K2-233 is the only young system with two detected inner planets below the radius valley and a third outer planet above it. This makes it an excellent target to perform comparative studies, to inform our theories of planet evolution, formation, migration, and atmospheric evolution.
\end{abstract}




\section{Introduction}

Young exoplanets ($<1$\,Gyr) are crucial to understanding planetary evolution. 
However, their detection is difficult with indirect methods, such as transit and radial velocity (RV), due to strong stellar signals imprinted in photometric and spectroscopic time-series. 
Significant progress has been made in the detection of transit signals in young stars' light curves. The time-scale difference between the transit and stellar signals in photometric data have enabled detections of tens of young exoplanets by filtering the relatively low-frequency stellar signals.
The \ktwo\ \citep[][]{Howell2014} and Transiting Exoplanet Survey Satellite \citep[\tess;][]{Ricker2015} missions
have discovered around a dozen transiting exoplanets or exoplanet candidates around young stars \citep[e.g.,][]{Bouma2020,David2018,Hobson2021,Martioli2021,Mann2016b,Mann2016a,Mann2021,Newton2019,Newton2021,Rizutto2020}.
A natural next step after finding a young transiting planet is to follow it up with spectroscopic observations in order to detect the RV wobbles induced by the planet on its host star. 
However, disentangling planet- and star-induced signals in RV data is still a big challenge in exoplanet science. 

Gaussian Processes (GPs) have became a popular mathematical framework to model activity-induced RVs given their flexibility to describe stochastic variations \citep[see e.g.,][]{Aigrain2022}. 
\citet[][]{Haywood2014} were the first to show that activity-induced RV signals can be modelled as a GP with a quasi-periodic covariance. 
Since then, multiple authors have used GPs to model stellar and planetary signals of active stars \citep[e.g.,][]{Dai2019,radvel,Grunblatt2015,Suarez2022}.
This approach creates flexible models for the RV time-series, where there are risks that potential planetary signals can be absorbed or modified by the GP activity model \citep[see e.g., discussions in][]{Ahrer2021,Rajpaul2021}.

Stellar activity also manifests in other observables that come from the same spectra from which the RVs are measured. They can be obtained from deformations of chromospheric lines, intensity of the lines, amongst others \citep[e.g,][]{Boisse2009,Queloz2001,Wilson1968}. 
This generates simultaneous ancillary time-series that contain information about the stellar activity and for this reason they are called \emph{activity indicators}. 
They usually contain only information about the stellar signal, and not about the planets \citep[although note that signals associated with planets can occur in activity signals, see][for an example]{Klein2022}. 
Therefore, they can be used to set priors on the pattern of the stellar activity in the RV time-series.
\citet[][]{Rajpaul2015} created a framework that uses spectroscopic activity-indicators together with RVs in order to set better constraints on the activity-induced signal in the RV time-series. 
This is done by combining all time-series within a multidimensional GP (hereafter multi-GP) framework that exploits the correlations between them.
\citet{Barragan2019} showed that this multi-GP can lead to the detection of planetary signals tens of times smaller than the stellar signal. This approach has now been used to spectroscopically confirm a handful of transiting young planets \citep[e.g.,][]{Barragan2022,Mayo2019,Nardiello2022,Zicher2022}.
Furthermore, \citet[][]{pyaneti2} showed (with simulated data) that the multi-GP approach can lead to more precise planetary signal detection in comparison with RV-only GP regressions \citep[see also][]{Ahrer2021,Rajpaul2021}.

In this manuscript, we explore this further with real data for the \target\ system.
K2-233 is a young ($\sim 360$\,Myr) star hosting three transiting planets with orbital periods of $\sim$ 2.5, 7, and 24 d \citep[][]{David2018}. \target's main identifiers and properties appear in Table~\ref{tab:parstellar}.
\citet[][]{Lillo2020} performed an intensive spectroscopic follow-up of the system, confirming the planetary nature of the three transiting signals and putting constraints on the composition of the three planets. In this work, we reanalyse the \citet[][]{Lillo2020} data sets with a multi-GP approach to put new constraints on the planetary masses.
This manuscript is part of a series of papers under the project 
\emph{GPRV: Overcoming stellar activity in radial velocity planet searches} funded by the European Research Council (ERC, P.I.~S.~Aigrain).
The outline of this manuscript is as follows: Section~\ref{sec:data} describes the \target\ photometric and spectroscopic data used for the analyses presented in Sect.~\ref{sec:datanalaysis}. Section~\ref{sec:discusion} is devoted to the discussion, followed by Sect.~\ref{sec:conclusions} where we summarise the most important results of this work.

\begin{table}
\caption{Main identifiers and parameters for \target.  \label{tab:parstellar} 
}
\begin{center}
\begin{tabular}{lcc} 
\hline
\noalign{\smallskip}
Parameter & Value &  Source \\
\noalign{\smallskip}
\hline
\noalign{\smallskip}
\multicolumn{3}{l}{\emph{Main identifiers}} \\
\noalign{\smallskip}
Gaia DR3  & 6253186686054822784 & \citet{Gaia2020}  \\
TYC & 6179-186-1 & \citet[][]{Hog2000} \\
2MASS & J15215519-2013539 &  \citet[][]{Cutri2003} \\
Spectral type & K3 & \citet{David2018} \\
\noalign{\smallskip}
\hline
\noalign{\smallskip}
\multicolumn{3}{l}{\emph{Equatorial coordinates, proper motion, and parallax}} \\
\noalign{\smallskip}
$\alpha$(J2000.0) &  15 21 55.1983 &  \citet{Gaia2020} \\
$\delta$(J2000.0) & -20 13 53.9909  & \citet{Gaia2020} \\
$\mu_\alpha$\,(mas\,${\rm yr^{{-1}}}$) & $ 	-20.031 \pm 0.052$ &  \citet{Gaia2020}  \\
$\mu_\delta$\,(mas\,${\rm yr^{{-1}}}$) & $ -30.963 \pm 0.020$ &   \citet{Gaia2020} \\
$\pi$\,(mas) & $ 	14.7719 \pm 0.0188$ & \ \citet{Gaia2020}  \\
Distance\,(pc) & $67.695 \pm 0.086$ & \citet{Gaia2020} \\
\noalign{\smallskip}
\hline
\noalign{\smallskip}
\multicolumn{3}{l}{\emph{Magnitudes}} \\
B &	$ 11.81 \pm 0.15$ &	 \citet[][]{Hog2000} \\  		
V &	$10.88 \pm 0.09$ &	  \citet[][]{Hog2000} \\ 
Gaia & $10.4229 \pm 0.0028 $ &	 \citet{Gaia2020}  \\ 	  		
J & $8.968 \pm 0.020$ &
	\citet{Cutri2003} 	 \\ 	  		
H &	$8.501 \pm 0.026$ &
	\citet{Cutri2003} 	 \\ 	  		
Ks &	$ 8.375 \pm 0.023$ &
	\citet{Cutri2003} 	 \\ 	  		
\noalign{\smallskip}
\hline
\noalign{\smallskip}
\multicolumn{3}{l}{\emph{Stellar parameters}} \\
$T_{\rm eff}$ (K) & $4796 \pm 66$ & \citet[][]{Lillo2020} \\
\logg (cgs) & $4.53 \pm 0.22$ & \citet[][]{Lillo2020} \\
$[Fe/H]$ & $-0.082 \pm 0.028$ & \citet[][]{Lillo2020} \\
Age (Myr) & $360_{-140}^{+490}$ & \citet[][]{David2018} \\
$P_{\rm rot}$ (d) & $9.754 \pm 0.038$ & \citet[][]{David2018} \\
\vsini\ (\kms) & $4.5 \pm 1.0$ & \citet[][]{David2018} \\
Mass ($M_\odot$) & $0.79 \pm 0.01$ & \citet[][]{Lillo2020} \\
Radius ($R_\odot$) & $0.71 \pm 0.01$ & \citet[][]{Lillo2020}\\
Density (\gcm) & $3.17 \pm 0.07 $ & \citet[][]{Lillo2020}\\
\hline
\end{tabular}
\end{center}
\end{table}

\section{\target\ data}
\label{sec:data}

\subsection{Light curve}
\label{sec:k2data}

\begin{figure*}
    \centering
    \includegraphics[width=\textwidth]{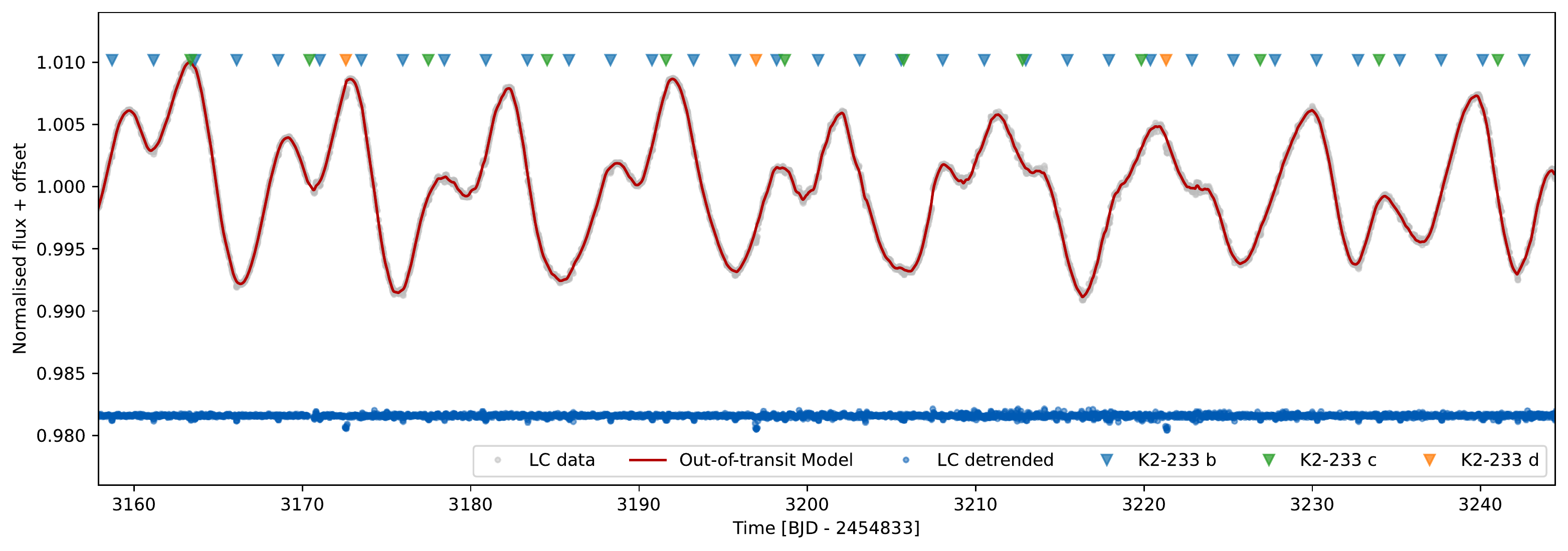} \\
    \caption{\ktwo\ light curve for \target.
    \ktwo\ data are shown with grey points with the out-of-transit variability model over-plotted in red. 
    The resulting flattened light curve is presented with blue points.
    Transit positions are marked with blue, orange, and green triangles for \target\, b, c and d, respectively.
    }
    \label{fig:lc}
\end{figure*}

\target\ was observed by \ktwo\ during its Campaign 15 between 2017 Aug 23 and 2017 Nov 20. Using these data, \citet[][]{David2018} discovered and validated three transiting planets with periods of 2.5, 7, and 24 d, and radii of $1.3$, $1.3$, and $2.6\,R_\oplus$, respectively. 
We downloaded \target's light curve as provided by \citet[][]{Vanderburg2014}\footnote{\url{https://lweb.cfa.harvard.edu/~avanderb/k2.html}.}. Figure~\ref{fig:lc} shows \target's \ktwo\ photometric time-series. \target's astrophysical signal is a result of activity regions on the stellar surface and transiting signals.

In order to analyse the transiting signals, we proceed to detrend the light curve using the public code \href{https://github.com/oscaribv/citlalicue}{\texttt{citlalicue} \faGithub} \citep{pyaneti2}.
In short, \citlalicue\ uses a GP regression
as implemented in \texttt{george} \citep[][]{george} to model the low frequency signals caused by stellar activity in the light curves.
We input to \citlalicue\ the light curves normalised to 1, as well as the ephemeris of the three transiting signals. We proceed by binning the data to 3 hours and masking out all the transits from the light curve while performing the GP regression using a Quasi-Periodic kernel {\citep[see][for more details]{george}}. 
We apply an iterative optimisation to find the maximum likelihood together with a $5$-sigma clipping algorithm to find the best model describing the out-of-transit variations. 
We then divide the \ktwo\ light curve by the inferred model to obtain a flattened light curve with transiting signals. 
Figure~\ref{fig:lc} shows the detrended light curve.

\subsection{Spectroscopic observations}

\citet[][]{Lillo2020} performed a spectroscopic follow up of \target\ using the High Accuracy Radial Velocity Planet Searcher \citep[HARPS;][]{Mayor2003} spectrograph mounted at the 3.6\,m ESO telescope at La Silla Observatory in Chile. They collected 124 HARPS spectra in a window of 105 d (Based on observations collected under ESO programmes 198.C-0169 and 095.C-0718).

The spectra obtained by \citet[][]{Lillo2020} are high quality with a typical S/N $> 50$  (at 550\,nm) per observation, which leads to a median RV uncertainty of 1.2\,\ms. 
However, we found that four observations have relatively large error bars, larger than 3\,\ms. They are the observations taken on 2458294.516674118 BJD (2018-06-25 00:24 UT), 2458301.658614401 BJD (2018-7-2 3:48 UT), 2458335.597976999 BJD (2018-8-5 2:21 UT), and 2458354.5192679004 BJD (2018-8-24 0:27 UT).  
We checked the observational conditions at La Silla when these data were collected using the ESO Astronomical Site Monitor (ASM) online tool\footnote{\url{https://www.eso.org/asm}.}.
We found that the observations taken on 2018-06-25 00:24 UT and 2018-8-24 0:27 UT were taken with bad seeing conditions ($> 2$).
In particular, the observation taken on 2018-06-25 00:24 UT suffers from a significant change of seeing during the acquisition, this generates an uneven light collection through the fibre that can create unwanted systematics effects in precision RV measurements.
We decided to remove the observations taken on 2018-06-25 00:24 UT and 2018-8-24 0:27 UT from our subsequent analyses.

We downloaded the \target's data taken from \citet[][]{Lillo2020} from the ESO archive.
We reduced the data using the dedicated HARPS data reduction software (DRS) and extracted the RV measurements by cross-correlating the Echelle spectra with a K5 numerical mask \citep{Baranne1996,Pepe2002, Lovis2007}. 
We also used the \texttt{DRS} to extract the Ca\,{\sc ii} H\,\&\,K lines (\sshk), \halpha, and \sodium\ chromospheric activity indicators, as well as two classic profile diagnostics of the cross-correlation function (CCF), namely, the full width at half maximum (FWHM), and the bisector inverse slope (BIS).
These reprocessed HARPS RV measurements have a typical error bar of 1.2\,\ms\ and an RMS of 9.1\,\ms. We note that these new RVs are equivalent to the values reported in \citet[][]{Lillo2020}, but this reprocessing of the spectra allowed us to extract tailored activity indicators. 
Table~\ref{tab:harps} lists the HARPS spectroscopic time-series. 

We acknowledge that there are eight archival HARPS observations of \target\ taken in 2014 (under the programs: 085.C-0019(A) by PI Lo Curto and 072.C-0488(E) by PI Mayor). We note that these observations have sub-optimal sampling in order to constrain the shape of the stellar signal in the spectroscopic time-series. Therefore, we do not include them in our analysis.

\begin{table*}
\begin{center}
\caption{HARPS spectroscopic measurements. The full version of this table is available in a machine-readable format as part of the supplementary material. \label{tab:harps}}
\begin{tabular}{ccccccccccc}
\hline\hline
Time & RV & $\sigma_{\rm RV}$ & FWHM & BIS & \sshk\ & $\sigma_{\rm S_{HK}}$ & \halpha\ & $\sigma_{\rm H_\alpha}$ & \sodium & $\sigma_{\rm Na}$ \\
${\rm BJD_{TDB}}$ - 2\,450\,000 & \kms & \kms & \kms & \kms &  &  \\
\hline
8257.575545 & -9.6220 & 0.0019 & 7.6508 & 0.0174 & 0.6893 & 0.0068 & 0.4774 & 0.0021 & 0.1713 & 0.0008 \\ 
8257.636216 & -9.6244 & 0.0020 & 7.6542 & 0.0156 & 0.6941 & 0.0073 & 0.4789 & 0.0022 & 0.1682 & 0.0009 \\ 
8257.755000 & -9.6260 & 0.0020 & 7.6635 & 0.0213 & 0.7034 & 0.0079 & 0.4802 & 0.0021 & 0.1693 & 0.0009 \\ 
8258.564438 & -9.6592 & 0.0019 & 7.6835 & 0.0566 & 0.7124 & 0.0065 & 0.4957 & 0.0021 & 0.1695 & 0.0008 \\ 
8258.658789 & -9.6663 & 0.0022 & 7.6591 & 0.0436 & 0.6915 & 0.0083 & 0.4868 & 0.0024 & 0.1682 & 0.0010 \\ 
$\cdots$ \\
\hline
\end{tabular}
\end{center}
\end{table*}

\section{Data analysis}
\label{sec:datanalaysis}

We performed a series of joint transit and spectroscopic time-series models. All our models are created using the same configuration for transit modelling (described in Section~\ref{sec:transitanalysis}). However, we change the way in which we modelled the spectroscopic time series in order to explore how different methods disentangle the planetary and stellar signals from the RV time-series (Sect.~\ref{sec:planetmasses}). 

For all the subsequent analyses we used the code \pyaneti\ \citep{pyaneti,pyaneti2}. In all our runs we sample the parameter space with 250 walkers using the Markov chain Monte Carlo (MCMC) ensemble sampler algorithm implemented in \pyaneti\ \citep{pyaneti,emcee}. 
The posterior distributions are created with the last 5000 iterations of converged chains. We thinned our chains by a factor of 10 giving a distribution of 125\,000  points for each sampled parameter. 

\subsection{Transit modelling configuration}
\label{sec:transitanalysis}

In order to perform the transit analysis we use the flattened \ktwo\ light curve (see Sect.~\ref{sec:k2data}). To speed up the transit modelling, we only model data chunks spanning a maximum of 4 hours on either side of each transit mid-time.
Since \ktwo\ Campaign 15 data were taken with a cadence of 30 min, we re-sampled the model over thirty steps to account for the data integration \citep{Kipping2010}.
We model the transits for the three planets, where for each planet we sample for the time of transit, $T_0$; orbital period, $P_{\rm orb}$; orbital eccentricity, $e$; angle of periastron, $\omega$; and scaled planetary radius $R_{\rm p}/R_\star$. We note that we also need to solve for the scaled semi-major axis ($a/R_\star$) for each planet. To do this we sample for the stellar density, $\rho_\star$, and we recover $a/R_\star$ for each planet using Kepler's third law \citep[see e.g.,][]{Winn2010}. 
We account for the limb darkening using the quadratic limb darkening approach described in \citet{Mandel2002} with the $q_1$ and $q_2$ parametrisation given by \citet{Kipping2013}.
We also sample for a photometric jitter term ($\sigma_{K2}$) to penalise the imperfections of our transit model.

Now we will describe the priors adopted for the transit parameters for all our runs. We adopt a beta distribution prior for the eccentricity, $\mathcal{B}(1.52,29)$, as advocated by \citet[][]{VanEylen2019} for multi-planet systems. For the angle of periastron we set a uniform prior ranging between $0$ and $2\pi$.
We set a Gaussian prior on the stellar density based on the stellar parameters given in Table~\ref{tab:parstellar}.
For the rest of the parameters, we set informative uniform priors.

\subsection{Detecting the planetary Doppler signals}
\label{sec:planetmasses}

\subsubsection{One-dimensional RV GP regression}
\label{sec:1dgprvs}

We first run a one-dimensional GP modelling of the RVs. We note that this analysis is equivalent to the one presented in \citet[][]{Lillo2020}. We model the stellar signal using a GP whose covariance between two times $t_i$ and $t_j$ is given by
\begin{equation}
    \gamma_{\rm 1D,{i,j}} = A^2 \gamma_{{\rm QP},i,j},
\end{equation}
\noindent
where $A$ is an amplitude term, and $\gamma_{{\rm QP},i,j}$ is the Quasi-Periodic (QP) kernel given by
\begin{equation}
   \gamma_{{\rm QP},i,j} = \exp 
    \left[
    - \frac{\sin^2[\pi(t_i - t_j)/P_{\rm GP}]}{2 \lambda_{\rm P}^2}
    - \frac{(t_i - t_j)^2}{2\lambda_{\rm e}^2}
    \right],
    \label{eq:gamma}
\end{equation}
\noindent
whose hyperparameters are, \pgp, the GP characteristic period; \lbp, the inverse of the harmonic complexity; and \lbe, the long-term evolution timescale. 

The planetary signals are included as a parametric mean function of the GP. We use 3 Keplerian signals, where every Keplerian signal is modelled with a time of minimum conjunction (equivalent to the time of transit for transiting planets), $T_0$; orbital period, $P_{\rm orb}$; orbital eccentricity, $e$; angle of periastron, $\omega$; and Doppler semi-amplitude, $K$. We also include one offset to account for the systemic velocity of the star, and a jitter term to penalise the imperfections of our model. 

We set wide uniform priors for all the GP hyperparameters. 
For \lbe\ we set a uniform prior between 0 and 500 d, for \lbp\ between 0.1 and 10, and for \pgp\ between 8 and 12 d.
We note that the Keplerian models share $T_0$, $P_{\rm orb}$, $e$, and $\omega$ with the transit modelling. Therefore, the priors for these parameters are described in Sect.~\ref{sec:transitanalysis}.
For the rest of the Keplerian parameters and the offset, we set uniform priors. 
For the jitter term, we set modified Jeffreys priors as defined by \citet{Gregory2005}.
The posterior creation follows the MCMC guidelines given at the beginning of Sect.~\ref{sec:datanalaysis}.

\begin{figure}
    \centering
    \includegraphics[width=0.48\textwidth]{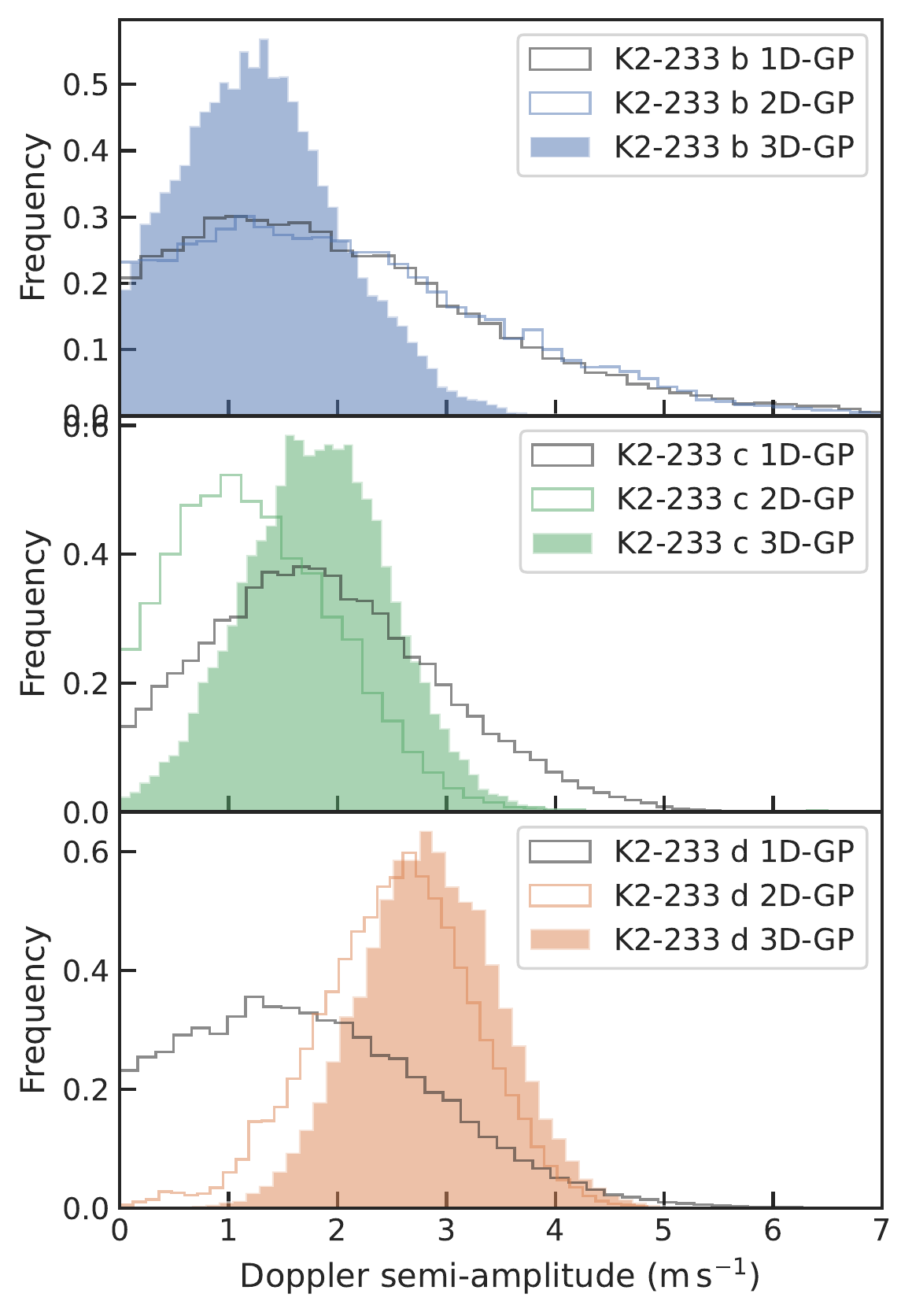}
    \caption{Posterior distributions for the Doppler semi-amplitude of \targetb\ (top), c (centre), and d (bottom). Shape of the posterior distributions for the one-(solid black line), two-(solid colourful line) and three-dimensional (filled colourful) GP analyses are shown.}
    \label{fig:kas}
\end{figure}

The recovered QP kernel hyperparameters are \lbe$=17.1 \pm 1.7$\,d, \lbp$=0.29 \pm 0.03$, and \pgp$=9.65 \pm 0.09$\,d. For a discussion on the interpretation of these hyperparameters we refer the reader to Appendix~\ref{sec:stellarsignal}.
Figure~\ref{fig:kas} shows the recovered posterior distributions for the Doppler semi-amplitudes for \targetbcd\ (black line histograms). 
The recovered values for the Doppler semi-amplitudes are $1.9_{ - 1.2 }^{+1.6}$\,\ms, $1.8 _{ - 1.0 }^{+1.1 }$\,\ms, and  $1.7_{ - 1.0 }^{ + 1.3 }$\,\ms for \targetbcd, respectively.
These values are consistent with the published values in \citet[][]{Lillo2020}. This is expected given that we are reproducing the same analysis in order to compare it with the analysis presented in Sect.~\ref{sec:multigps}. All these values are constrained to less than a $2-$sigma confidence level.
We note that exploring the posteriors for the orbital eccentricities they all are consistent with a zero eccentricity and the shape of the posterior follows the beta prior.

\subsubsection{Multidimensional GP regression}
\label{sec:multigps}

We then perform a multi-GP approach to characterise the stellar and planetary signals in our RV time-series \citep[see][for more details]{Rajpaul2015,pyaneti2}.
We create $N-$dimensional GP models, including $N$ time-series $\mathcal{A}_i$,  as
\begin{equation}
\begin{matrix}
 \mathcal{A}_1 =  A_{1} G(t) + B_{1} \dot{G}(t) \\
\vdots \\
\mathcal{A}_N =  A_{N} G(t) + B_{N} \dot{G}(t), \\
\end{matrix}
\label{eq:gps}
\end{equation}
\noindent
where the variables $A_{1}$, $B_{1}$, $\cdots$, $A_{N}$, $B_{N}$, are free parameters which relate the individual time-series to $G(t)$ and $\dot{G}(t)$. In this approach, $G(t)$ is a latent (unobserved) variable, which can be loosely interpreted as representing the projected area of the visible stellar disc that is covered in active regions as a function of time. 

We perform a 2-dimensional GP model between the RVs and FWHM, and a 3-dimensional run between RVs, FWHM, and BIS. For a further discussion on the use of other activity indicators we encourage the reader to check Appendix~\ref{sec:stellarsignal}.
The multidimensional covariance matrix was created using the QP kernel given in eq.~\eqref{eq:gamma} and its derivatives as described in \citet[][]{Rajpaul2015} and \citet[][]{pyaneti2}. 
We assume that RVs and BIS time-series can be described as $\mathcal{A}_i =  A_{i} G(t) + B_{i} \dot{G}(t)$, while the FWHM time-series is photometric-like only, i.e., it is described as $\mathcal{A}_i =  A_{i} G(t)$.
For all runs, the mean function corresponding to the RVs was created using three Keplerian curves following the same setup described in Sect.~\ref{sec:1dgprvs}, including the offset and jitter terms. We recall the reader that these runs are done together with the transit analysis following the setup described in Sect.~\ref{sec:transitanalysis}.
For the activity indicators, the mean function was treated as an offset. We also include a jitter term for each activity indicator.
The priors for the mean functions and QP kernel hyperparameters have the same ranges as described in Sect.\ref{sec:1dgprvs}. 
We create posterior distributions using the same MCMC approach as in Sect.~\ref{sec:transitanalysis}.

For the 2-dimensional GP analysis of the RVs and FWHM, we recover the hyperparameters \lbe$=16.2 \pm 2.0$\,d, \lbp$=0.46 \pm 0.05$, and \pgp$=9.73 \pm 0.08$\,d. 
For a more detailed discussion on the interpretation of the hyperparameters we refer the reader to Appendix~\ref{sec:1dgprvs} and to \citet[][]{Nicholson2022}.
The recovered values for the Doppler semi-amplitudes are $1.9_{ - 1.2 }^{+1.6}$\,\ms, $1.18 _{ - 0.69}^{+0.86 }$\,\ms, and $2.57_{ - 0.74 }^{ + 0.68 }$\,\ms\ for \targetbcd, respectively.
Figure~\ref{fig:kas} shows the recovered posterior distributions for the recovered Doppler semi-amplitude for the three planets (coloured, unshaded histograms).
This analysis gives a detection of \targetd\ with a significance of detection of $3.5-$sigma. While for planet b the posterior does not change significantly, and there is an improvement in the detection of planet c, but still below the $2-$sigma significance.
This improvement on the detection of the planet signal can be explained by a better constraint of the stellar activity by using the information of the stellar signal in the FWHM time-series.
It is worth noting that the posterior distribution for the eccentricity of planet d does not follow the prior. 
The inferred eccentricity of planet d is $e_d = 0.12_{-0.07}^{+0.09}$, despite the prior pushing for smaller values. 

For the three-dimensional GP analysis of the RVs, FWHM, and BIS, we recover the hyperparameters \lbe$=18.0 \pm 2.1$\,d, \lbp$=0.50 \pm 0.05$, and \pgp$=9.75 \pm 0.07$\,d. 
For this new analysis, the recovered values for the Doppler semi-amplitudes for \targetbcd\ are $1.26_{ - 0.71 }^{+0.79}$\,\ms, $1.82 _{ - 0.69 }^{+0.67 }$\,\ms, and $2.83 \pm 0.65$\,\ms, respectively. As in the two-dimensional GP regression, we also have an improvement on the RV Doppler detection of the three planets, in this case, we detect \targetbcd\ with a significance of $1.8-$, $2.7-$ and $4.3$-sigma, respectively.  Figure~\ref{fig:kas} shows the Doppler semi-amplitude posterior distributions coming from the three-dimensional GP analysis (coloured, shaded histograms).
We can see that the inclusion of the BIS time-series does not lead to a significant improvement on the detection of \targetd. However, this three-dimensional analysis does boost the detection of \targetc. In this particular case, the inclusion of the BIS time-series helps to constrain the activity signal and disentangle it from the Doppler signal of one of the planets. 
We note too that, although we do not quite achieve a 3-$\sigma$ detection of \targetb's Doppler signal, the range of possible masses is better constrained than before. The derived mass for this planet is \mpb\ with a maximum mass of $6.12\,M_\oplus$ at the 99\% confidence level.
Similarly to the two-dimensional GP analysis, we found that the posterior distribution for \targetd's eccentricity does not follow the prior and in this case, the inferred value is $e_d = 0.19 \pm 0.09$.

\subsection{Final model}
\label{sec:final}

Based on the analyses presented in this section, our final model for the photometric and spectroscopic data of \target\ is the transit model described in Section~\ref{sec:transitanalysis}, together with the three-dimensional GP regression described in Sect.~\ref{sec:multigps} (In Sect.~\ref{sec:comparison} we describe the reasoning to prefer this model over the others). 
For the RV planetary model, we assume that the two innermost planets have circular orbits while allowing for an eccentricity for \targetd. The whole set of sampled parameters and priors are shown in Table~\ref{tab:pars}.

\begin{figure*}
    \centering
    \includegraphics[width=\textwidth]{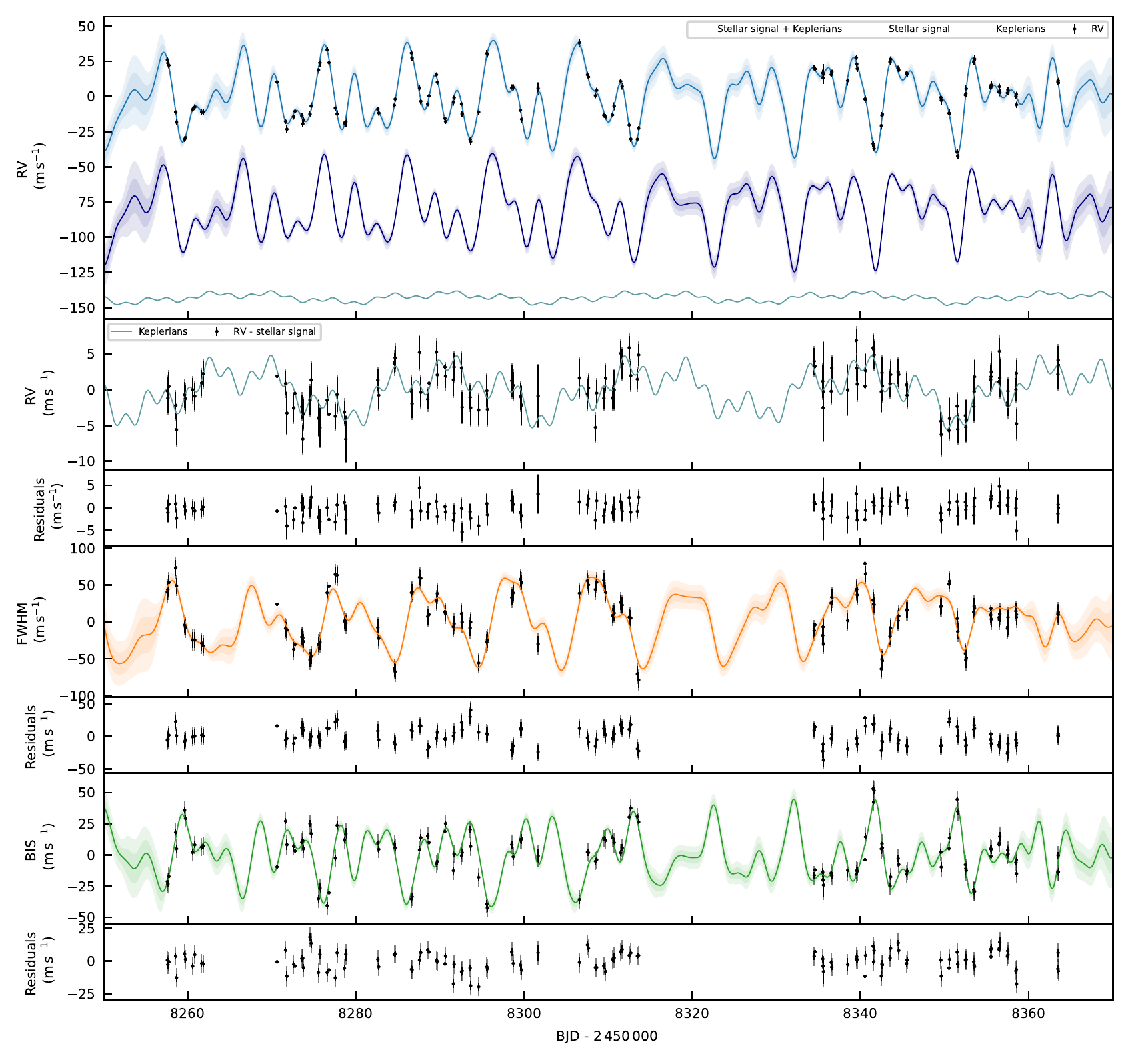}
    \caption{\target's RV, FWHM and BIS time-series after being corrected by inferred offsets. The plot shows (from top to bottom): RV data together with full, stellar and planetary signal inferred models; RV data with stellar signal model subtracted; RV residuals; FWHM data together with inferred stellar model; FWHM residuals; BIS data together with inferred stellar model, and BIS residuals.
    Measurements are shown with black circles, error bars, and a semi-transparent error bar extension accounting for the inferred jitter. 
    The solid lines show the inferred full model coming from our multi-GP, light-shaded areas showing the corresponding GP model's one and two-sigma credible intervals.
    For the RV time-series (top panel) we also show the inferred stellar (dark blue line) and planetary (light green line) recovered signals with an offset for better clarity.}
    \label{fig:timeseries}
\end{figure*}

\begin{figure*}
    \centering
    \includegraphics[width=\textwidth]{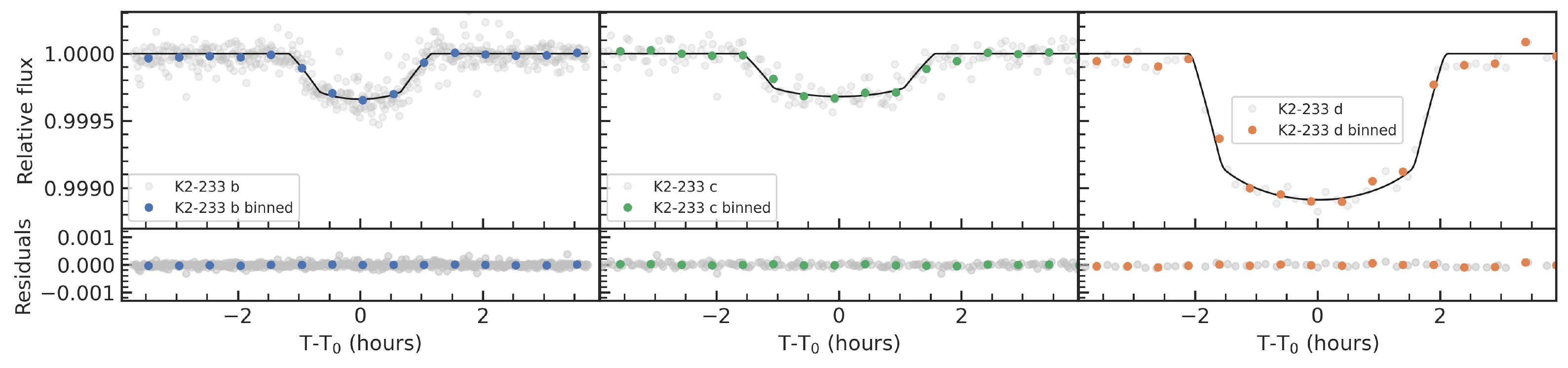}
    \includegraphics[width=0.98\textwidth]{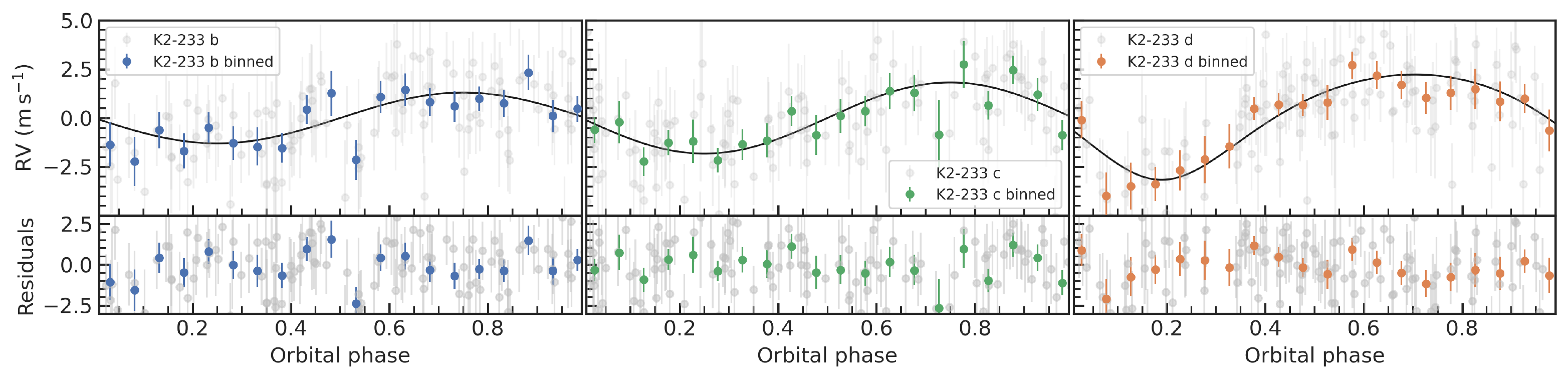}
   \caption{
    \emph{Top panel}: Phase-folded light curves of \target\,b (left), \target\,c (centre) and \target\,d (right). Nominal \ktwo\ observations are shown in light grey. Solid colour circles represent 30-min binned data. Transit models are shown with a solid black line. The x- and y-axis in each panel are shown with the same range to facilitate the comparison of the scales of the transit signals.
   \emph{Bottom panel}: Phase-folded RV signals for \targetb\ (left), \targetc\ (centre), and \targetd\ (right) following the subtraction of the systemic velocities, stellar signal, and other planets. Nominal RV observations are shown as light grey points. Solid colourful points show binned data to 1/20 of the orbital phase.
    }
    \label{fig:rvfolded}
\end{figure*}

Figure~\ref{fig:timeseries} shows the spectroscopic time-series plots resulting from this joint analysis, while Fig.~\ref{fig:rvfolded} shows the independent transit and Doppler signals for each of the planets.
Table~\ref{tab:pars} shows the inferred sampled parameters, defined as the median and 68.3\% credible interval of the posterior distribution.
Table~\ref{tab:derived} shows all the derived planetary and orbital parameters.
The detected Doppler semi-amplitudes \kb[], \kc[], and \kd, translate to masses of \mpb[], \mpc[], and \mpd\ for planets \targetbcd, respectively. 
This new methodology provides higher precision masses than those reported in the original detection and confirmation work by \citet[]{Lillo2020}.
Therefore, we will reassess the implications on planetary composition and perspectives on atmospheric characterisation in Sect.~\ref{sec:discusion}.

\begin{table*}
\begin{center}
  \caption{Model parameters and priors for joint fit \label{tab:pars}}  
  \begin{tabular}{lcc}
  \hline
  \hline
  \noalign{\smallskip}
  Parameter & Prior$^{(a)}$ & Final value$^{(b)}$ \\
  \noalign{\smallskip}
  \hline
  \noalign{\smallskip}
  \multicolumn{3}{l}{\emph{\bf \targetb's parameters }} \\
  \noalign{\smallskip}
    Orbital period $P_{\mathrm{orb}}$ (days)  & $\mathcal{U}[2.4667 , 2.4683 ]$ &\Pb[] \\
    Transit epoch $T_0$ (BJD$_\mathrm{TDB}-$2\,450\,000)  & $\mathcal{U}[7991.6745 , 7991.7077]$ & \Tzerob[]  \\  
    Scaled planet radius $R_\mathrm{p}/R_{\star}$  &$\mathcal{U}[0.0,0.05]$ & \rrb[]  \\
    Impact parameter, $b$ & $\mathcal{U}[0,1]$ & \bb[] \\
    Orbital eccentricity, $e$ & $\mathcal{F}[0]$ & 0 \\
    Angle of periastron, $\omega$(deg) & $\mathcal{F}[90]$ & $90$ \\
    Doppler semi-amplitude variation $K$ (m s$^{-1}$) & $\mathcal{U}[0,50]$ & \kb[] \\
  \multicolumn{3}{l}{\emph{\bf \targetc's parameters }} \\
    Orbital period $P_{\mathrm{orb}}$ (days)  & $\mathcal{U}[7.0547 , 7.0655]$ &\Pc[] \\
    Transit epoch $T_0$ (BJD$_\mathrm{TDB}-$2\,450\,000)  & $\mathcal{U}[7586.8425 , 7586.9105]$ & \Tzeroc[]  \\  
    Scaled planet radius $R_\mathrm{p}/R_{\star}$  &$\mathcal{U}[0.0,0.05]$ & \rrc[]  \\
    Impact parameter, $b$ & $\mathcal{U}[0,1]$ & \bc[] \\
    Orbital eccentricity, $e$ & $\mathcal{F}[0]$ & 0 \\
    Angle of periastron, $\omega$(deg) & $\mathcal{F}[90]$ & $90$ \\
    Doppler semi-amplitude variation $K$ (m s$^{-1}$) & $\mathcal{U}[0,50]$ & \kc[] \\
  \multicolumn{3}{l}{\emph{\bf \targetd's parameters }} \\
    Orbital period $P_{\mathrm{orb}}$ (days)  & $\mathcal{U}[24.3509 , 24.3781 ]$ &\Pd[] \\
    Transit epoch $T_0$ (BJD$_\mathrm{TDB}-$2\,450\,000)  & $\mathcal{U}[8005.5640 , 8005.6008 ]$ & \Tzerod[]  \\  
    Scaled planet radius $R_\mathrm{p}/R_{\star}$  &$\mathcal{U}[0.0,0.05]$ & \rrd[]  \\
    Impact parameter, $b$ & $\mathcal{U}[0,1]$ & \bd[] \\
    Orbital eccentricity, $e$ & $\mathcal{B}[1.52,29]^{(\mathrm{c})}$ & \ed[] \\
    Angle of periastron, $\omega$ (deg) & $\mathcal{U}[0,360]$ & \wdd[] \\
    Doppler semi-amplitude variation $K$ (m s$^{-1}$) & $\mathcal{U}[0,50]$ & \kd[] \\
    \multicolumn{3}{l}{\emph{ \bf GP hyperparameters}} \\
   GP Period $P_{\rm GP}$ (days) &  $\mathcal{U}[8,12]$ & \jPGP[] \\
    $\lambda_{\rm p}$ &  $\mathcal{U}[0.1,5]$ &  \jlambdap[] \\
    $\lambda_{\rm e}$ (days) &  $\mathcal{U}[1,500]$ &  \jlambdae[] \\
    $A_{\rm RV}$ (\ms)  &  $\mathcal{U}[0,100]$ & \jAzero \\
    $B_{\rm RV}$ (\ms\,d) &  $\mathcal{U}[-100,100]$ & \jAone \\
    $A_{\rm FWHM}$ (\ms) &  $\mathcal{U}[0,100]$ & \jAtwo \\
    $B_{\rm FWHM}$ (\ms\,d) &  $\mathcal{F}[0]$ & 0 \\
    $A_{\rm BIS}$ (\ms) &  $\mathcal{U}[-100,100]$ & \jAfour \\
    $B_{\rm BIS}$ (\ms\,d) &  $\mathcal{U}[-100,100]$ & \jAfive \\
    \multicolumn{3}{l}{\emph{ \bf Other parameters}} \\
    Stellar density $\rho_\star$ (\gcm) & $\mathcal{N}[3.17, 0.14]$ & \denstrb[] \\ 
    \ktwo\ Parameterised limb-darkening coefficient $q_1$  &$\mathcal{U}[0,1]$ & \qone \\ 
    \ktwo\ Parameterised limb-darkening coefficient $q_2$  &$\mathcal{U}[0,1]$ & \qtwo \\ 
    Offset RV (\kms) & $\mathcal{U}[ -10 , -9]$ & \RV[] \\
    Offset ${\rm FWHM}$  (\kms)& $\mathcal{U}[ 7, 8]$ & \FWHM[]  \\
    Offset ${\rm BIS}$  (\kms)& $\mathcal{U}[ -0.5 , 0.5]$ & \BIS[]  \\
    Jitter term $\sigma_{\rm RV}$ (\ms) & $\mathcal{J}[1,100]$ & \jRV[] \\
    Jitter term $\sigma_{\rm FWHM}$ (\ms) & $\mathcal{J}[1,100]$ & \jFWHM[] \\
    Jitter term $\sigma_{\rm BIS}$ (\ms) & $\mathcal{J}[1,100]$ & \jBIS[] \\
    Jitter term $\sigma_{K2}$ (ppm) & $\mathcal{J}[1,100]$ & \jtr[] \\
    \noalign{\smallskip}
    \hline
\multicolumn{3}{l}{\footnotesize $^a$ $\mathcal{F}[a]$ refers to a fixed value $a$, $\mathcal{U}[a,b]$ to an uniform prior between $a$ and $b$, $\mathcal{N}[a,b]$ to a Gaussian prior with mean $a$ and standard deviation $b$,}\\
\multicolumn{3}{l}{  $\mathcal{B}[a,b]$ to a beta distribution with shape parameters $a$ and $b$, and $\mathcal{J}[a,b]$ to the modified Jeffrey's prior as defined by \citet[eq.~16]{Gregory2005}.}\\
\multicolumn{3}{l}{\footnotesize $^b$ Inferred parameters and errors are defined as the median and 68.3\% credible interval of the posterior distribution.}\\
\multicolumn{3}{l}{\footnotesize $^c$ Beta distribution to inform eccentricity sampling using the beta distribution for multi planetary systems as defined by \citet[][]{VanEylen2019}.}
  \end{tabular}
\end{center}
\end{table*}

\begin{table*}
\begin{center}
  \caption{Derived parameters for the \target\ planets. \label{tab:derived}}
  \begin{tabular}{lccc}
  \hline
  \hline
  Parameter & \targetb's  & \targetc's & \targetd \\
   & values & values & values \\
  \hline
  \noalign{\smallskip}
    Planet mass $M_\mathrm{p}$ ($M_{\rm \oplus}$) &  \mpb[]  &  \mpc[]  &  \mpd[] \\
    Planet radius $R_\mathrm{p}$ ($R_{\rm \oplus}$) &  \rpb[] &  \rpc[] & \rpd[] \\
    Planet density $\rho_{\rm p}$ (g\,cm$^{-3}$) &  \denpb[] &  \denpc[] & \denpd[] \\
    Scaled semi-major axis  $a/R_\star$ &  \arb[]  &  \arc[]  & \ard[] \\
    Semi-major axis  $a$ (AU) &  \ab[] &  \ac[] & \ad[] \\
    Time of periastron passage $T_p$ (BJD-2450000) & \Tzerob[] & \Tzeroc[] & \Tperid[] \\
    Orbit inclination $i_\mathrm{p}$ ($^{\circ}$) &  \ib[] &  \ic[] & \id[] \\
    Transit duration $\tau_{14}$ (hours) & \ttotb[]  & \ttotc[] & \ttotd[] \\
    Planet surface gravity $g_{\rm p}$ (${\rm cm\,s^{-2}}$)$^{(a)}$ & \grapb[] & \grapc[] & \grapd[] \\
    Planet surface gravity $g_{\rm p}$ (${\rm cm\,s^{-2}}$)$^{(b)}$ & \grapparsb[] & \grapparsc[] & \grapparsd[] \\
    Equilibrium temperature  $T_\mathrm{eq}$ (K)$^{(c)}$  &   \Teqb[] &   \Teqc[] & \Teqd[] \\
    Received irradiance ($F_\oplus$) & \insolationb[] & \insolationc[] & \insolationd[] \\
    TSM$^{(d)}$ & \tsmb[] & \tsmc[] & \tsmd[] \\
   \noalign{\smallskip}
  \hline
  \multicolumn{3}{l}{$^a$ Derived using $g_{\rm p} = G M_{\rm p} R_{\rm p}^{-2}$.}\\
  \multicolumn{3}{l}{$^b$ Derived using sampled parameters following \citet{Sotuhworth2007}.}\\
  \multicolumn{3}{l}{$^c$ Assuming a zero albedo.}\\
  \multicolumn{3}{l}{$^d$ Transmission spectroscopy metric (TSM) by \citet{Kempton2018}.}\\
  \end{tabular}
\end{center}
\end{table*}

\begin{figure}
    \centering
    \includegraphics[width=0.5\textwidth]{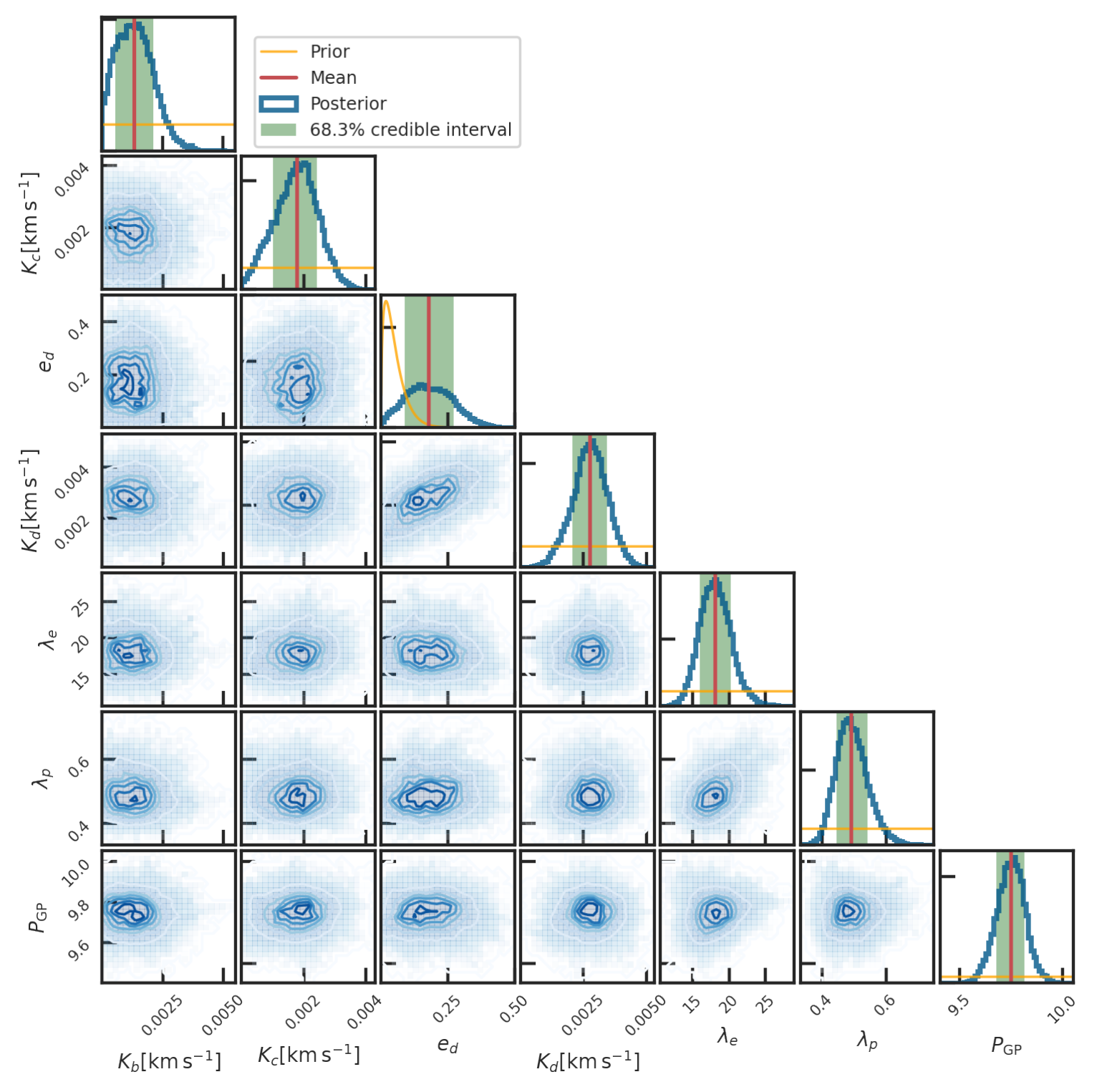}
    \caption{Posterior and correlation plots for some of the {key} sampled parameters of the joint analysis described in Sect.~\ref{sec:final}.}
    \label{fig:correlations}
\end{figure}

\section{Discussion}
\label{sec:discusion}

\subsection{Comparison between models}
\label{sec:comparison}

{
In sect.~\ref{sec:planetmasses} we have shown how the multi-GP analysis provides better constrained posterior distributions for the planetary-induced RV signals in comparison with the RV-only one-dimensional GP regression. 
From a qualitative point of view, a priori we know that there is an intrinsic correlation between the stellar signal in the RVs and activity indicators time-series \citep[see e.g,][]{Dumusque2014}. 
In the case where there is no instrumental red-noise in our time-series, we can assume that a model that uses activity indicators to constrain the stellar signal in the RV time-series is better than a model of the RVs only. 
This is especially important while doing GP regression, since an unconstrained GP predictive distribution can absorb or modify the signal from the mean function. 
On the other hand, the multi-GP approach sets a tighter constraint on the GP predictive distribution for the activity model in the RV time-series, which leads to less biased recovery of the mean model and, consequently, better inference of the mean function parameters \citep[see e.g,][for a more extensive discussion on this]{pyaneti2}. 
}

Given the Bayesian approach of the analyses, Bayesian Information Criterion or Bayesian evidence would be the best option to perform a quantitative comparison between our models. 
However, these methods are useful to compare different models to the same data set. Therefore, we cannot compare the one-dimensional GP model with the two- and three-dimensional GP models given that they all contain different time-series \citep[see also][for a discussion of this problem]{Ahrer2021,Rajpaul2021}.

\citet[][]{Rajpaul2021} proposed a solution to perform Bayesian comparisons for multi-GP regressions of different dimensions. 
Briefly, they modelled the activity indicators simultaneously with the RVs, but each time-series was modelled with different set of parameters that are not connected in any way. 
The activity indicators (in their case BIS and \logr) are modelled with a constant offset and jitter white noise. As it is argued in that manuscript, this approach has no effect on the inference of the RV-related planetary and orbital parameters. And  this approach also ensures that the model evidences were consistently normalised (having the same number of data), allowing comparisons between all of their models.
However, we notice that with this approach, the underlying model that describes the activity indicators is strongly penalised by default given that they are modelled only as an offset and a jitter term. 
Therefore, the difference in Bayesian evidence does not only reflect the model difference given by the planetary and activity model in the RVs, but it also is penalised by the sub-optimal modelling of the stellar signal in the activity indicators. We therefore assume that this approach is not optimal to compare models with different time-series. 

{
To show quantitatively the advantages on using the multi-GP, we show the different conclusions that we could draw for the different analyses regarding the RV detection of the planetary signals. 
The question we want to answer is: can we claim a RV detection of the planetary signals in each analysis? And using the Bayesian Information Criterion (BIC) and Akaike Information Criterion (AIC) we select the preferred model. 
We note that BIC and AIC have different  properties \citep[see e.g.,][]{burnham2002model,Chakrabarti2011}. BIC has been found more useful in helping to select the true model (assuming the true model is in the sample of tested models). While the AIC is more appropriate in finding the best model (assuming that we do not know the true model).
Since we do not believe that any of our models is the true one describing our data, AIC is the best metric to use. However, we will show BIC and AIC given the popularity of the former in the exoplanet comunity.  

For the onedimensional GP framework we create two models. One model in which we fix the Keplerian signals to have an amplitude of zero, and another model where we sample for them. 
Both models give similar likelihoods, but the former model is preferred with a $\Delta {\rm BIC} = 19$ and $\Delta {\rm AIC} = 4$. We therefore conclude that from the onedimensional GP regression we cannot conclude that we detect the RV planetary signals (even if a priori we know they should be there).

We repeat the same experiment but now within the threedimensional GP regression framework between RVS, FWHM, and BIS. 
In this case the model including the RV planetary signals is preferred over the model where they are zero with a $\Delta {\rm BIC} = 3$ and $\Delta {\rm AIC} = 19$. 
We thus conclude that with the multidimensional GP analysis we can claim, with a statistical significance, that the model including the planetary Doppler signals is preferred.

These analyses show how for the current dataset, the only method that allows to significantly claim the detection of the planetary RV signals is the multi-GP regression. 
This shows how the multi-GP analysis offers a more robust way of modelling spectroscopic time-series of young active stars and detecting the RV planetary signals in them.
}

\subsection{Dynamical analysis}

The architecture of \target\ is shaped for the orbits of \targetb\ and c being consistent with circular, and \targetd\ with a slight eccentricity of $\sim 0.18$. 
This relatively small eccentricity of planet d can be reached by planet-disc interaction \citep[e.g.,][]{Ragusa2018}. We then performed an orbital stability analysis of the \target\ system to test if this configuration of the system is stable. 
We did this using the software \texttt{mercury6} \citep{Chambers1999}. 
We assume that all planets have co-planar orbits and we use the median of the derived parameters in Tables~\ref{tab:pars} and \ref{tab:derived} to create our \texttt{mercury6} set-up. 
We evolved the system for 1 Gyr with steps of 0.1\,d per integration. 

We found negligible changes in the orbital parameters of the three planets. Except for the eccentricities that had fluctuations. \targetb\ and c presented eccentricity fluctuations of a maximum of 0.12. While \targetd's eccentricity fluctuations were in the 0.01 level. 
Therefore, we conclude that our orbital solution for the \target\ system is consistent with a dynamically stable configuration.

\subsection{Composition of the planets}

\begin{figure*}
    \centering
    \includegraphics[width=0.99\textwidth]{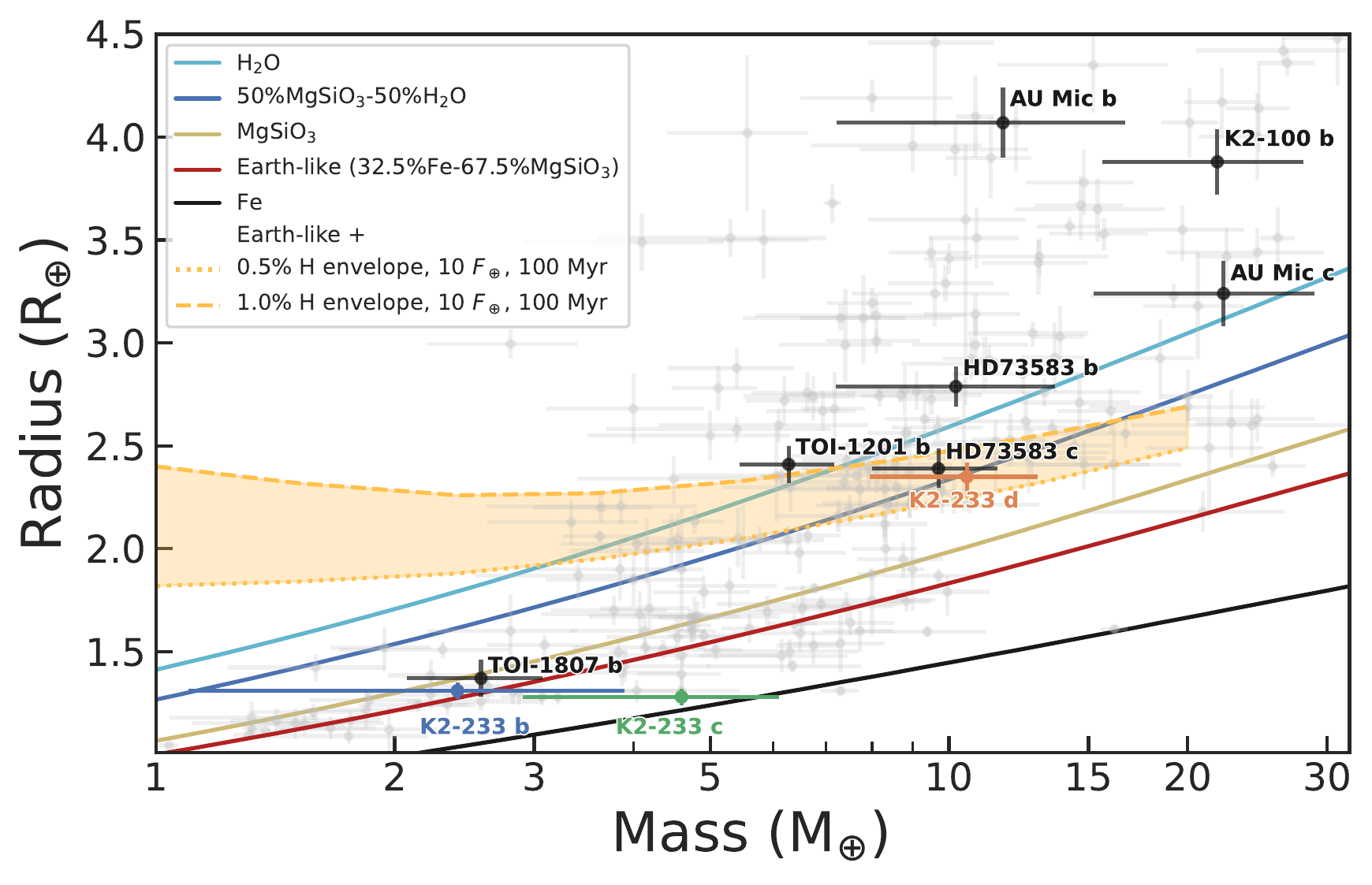}
    \caption{Mass \emph{vs} radius diagram for small exoplanets ($1 < R_{\rm p} < 4\, R_\oplus$ and $1 < M_{\rm p} < 32\, M_\oplus$). Grey points with error bars show planets with mass and radius measurements better than 30\% \citep[As in the NASA Exoplanet Archive on Feb 02, 2023, \url{https://exoplanetarchive.ipac.caltech.edu/}, ][]{NASAexoplanet}. 
    Black points and labels refer to young exoplanets ($< 1$\,Gyr) with mass and radius measurements. 
    \targetbcd\ are shown with blue, green, and orange circles with corresponding labels identifying each planet. Solid lines represent two-layer models as given by \citet{Zeng2016} with a different colour corresponding to a different mixture of elements. 
    Non-solid lines correspond to rocky cores surrounded by an Hydrogen envelope with 0.5\% (dotted line) and 1\% (dashed) on mass for 100\,Myr planets with an insolation of $10\,F_\oplus$ \citep{Lopez2014}.
    This plot was created using the same code used to create the mass-radius diagram in \citet[][]{Barragan2018b}.
    }
    \label{fig:mr}
\end{figure*}

Figure~\ref{fig:mr} shows a mass-radius diagram for small exoplanets ($1 < R_{\rm p} < 4.5\, R_\oplus$ and $1 < M_{\rm p} < 32 M_\oplus$). 
The plot also shows two layer exoplanet models by \citet{Zeng2016}, together with the Earth-like interior plus Hydrogen envelope models given by \citet{Lopez2014} for 100 Myr old planets receiving an insolation of $10\,F_\oplus$.

With a mass of \mpd\ and radius of \rpd, \targetd\  has a density of \denpd, placing it in the water world composition line regime (See Fig.~\ref{fig:mr}). 
Therefore, we can say that \targetd\ could be consistent with a water-rich world that consists of 50\% water ice with a small 50\% of silicates. However, \targetd\ is also consistent with a planet made of an Earth-like interior, surrounded by a Hydrogen envelope that could account for 0.5-1\% of its mass (taking into account its age and insolation). 
It is important to note that \targetd\ lies well above the small planet radius valley \citep[a lack of planets between $1.5$ and $2 R_\oplus$,][]{Fulton2017}. 
We therefore expect that the planet has a volatile envelope rather than being a solid water world \citep[as suggested by previous works e.g.,][]{Fulton2017,VanEylen2018}.
Despite the degeneracy in composition for \targetd\ inferred from the mass-radius diagram,  for the remainder of the discussion in this manuscript, we will assume that \targetd\ is a world with a Hydrogen-rich volatile envelope. 

The other two planets, \targetb\ and c, lie below the radius valley. Therefore, we would expect them to have a rocky composition \citep[as discussed by][]{Lillo2020}.
\citet{owenwu2017} posit that the position of the radius valley points to a universal Earth-like rock and iron core composition. Beyond that, it is interesting to compare the expected composition between these two innermost planets given their similar size. 
The measured mass and radius of \targetc\ present a relatively high density of \denpc. We can see in Figure~\ref{fig:mr} that, at face value, \targetc\ is consistent with a planet made principally of iron with a small fraction of silicates. This result is not unexpected, and it is consistent with the dispersion of exoplanets from that part of the mass-radius diagram. We must be cautious when interpreting our results for \targetb, given the marginal detection of the signal in the RV time-series. Its location in the mass-radius diagram suggests a lower density (hence more silicates and less iron) than \targetc, but the two planets' densities are actually consistent at the 1-$\sigma$ level. Therefore, we cannot claim to have detected a significant difference in composition between the two small planets in the system. 

Figure~\ref{fig:mr} also shows the position of other young small planets with mass and radius measurements. It is worth noting that \target\ is the only young system with characterised planets on both sides of the radius valley. 
Therefore, this system is a valuable laboratory for comparative studies of planetary evolution and formation, and to test theories such as photo-evaporation \citep[e.g.,][]{Adams2006,Lopez2014,Raymond2009,Owen2013} and core-powered mass-loss \citep[e.g.,][]{Ginzburg2016,Gupta2019,Gupta2021}.
For the rest of this discussion section, we will assume that \targetb\ and c are solid rocky worlds, while \targetd\ is a planet with a significant volatile envelope accounting for most of its radius.
Nonetheless, we note that further observations are needed in order to reduce the mass uncertainties and pinpoint the true composition of these worlds more precisely.

\subsection{Photo-evaporation and the radius valley}

The true origin of the small planet radius valley is still disputed. Gas-poor formation and a subsequent distinct, inherently rocky super-Earth population has mostly been ruled out, considering the slope of the radius valley \citep{VanEylen2018,petigura2022}. However, atmospheric erosion due to (i) photo-evaporation \citep{owenwu2017,VanEylen2018} and (ii) core-powered mass loss \citep{Ginzburg2016,Gupta2019,berger2023} are both theories which could explain the observed under-abundance of planets sized between $\sim 1.5$ and $2\,R_\oplus$. 

We note that \target\ has three, mutually aligned, close-in planets, therefore, we expect that they migrated through the disc, and hence have been on close orbits since $\lesssim 5$\,Myr and have been suffering from high levels of stellar radiation since then \citep[see e.g.,][]{Kley2012}. Since arriving at their present-day orbits, the planets will have been subjected to high levels of stellar irradiation. 
Theory predicts photo-evaporation to be more effective at driving hydrodynamic escape during the first $\sim100$\,Myr after formation \citep[e.g.,][]{owenwu2017}. 
At an age of 360\,Myr \citep[][]{David2018}, the K2-233 system is thus ideal to test the role of photo-evaporation in sculpting the radius valley.
On the other hand, we would not expect the planets in the K2-233 system to have suffered a significant atmospheric loss due to core-powered mass loss, since this mechanism operates over Gyr timescales \citep{Ginzburg2016}. 
The characteristics of the inner K2-233 system are suggestive of a history of significant atmospheric mass loss, which would support the photo-evaporation evolution scenario.
A more conclusive test, however, would be to search for signatures of ongoing evaporation, for example by searching for escaping Helium in transmission \citep[e.g.,][]{Zhang2022}.

\subsection{Atmospheric characterisation perspectives}

\begin{figure}
    \centering
    \includegraphics[width=0.48\textwidth]{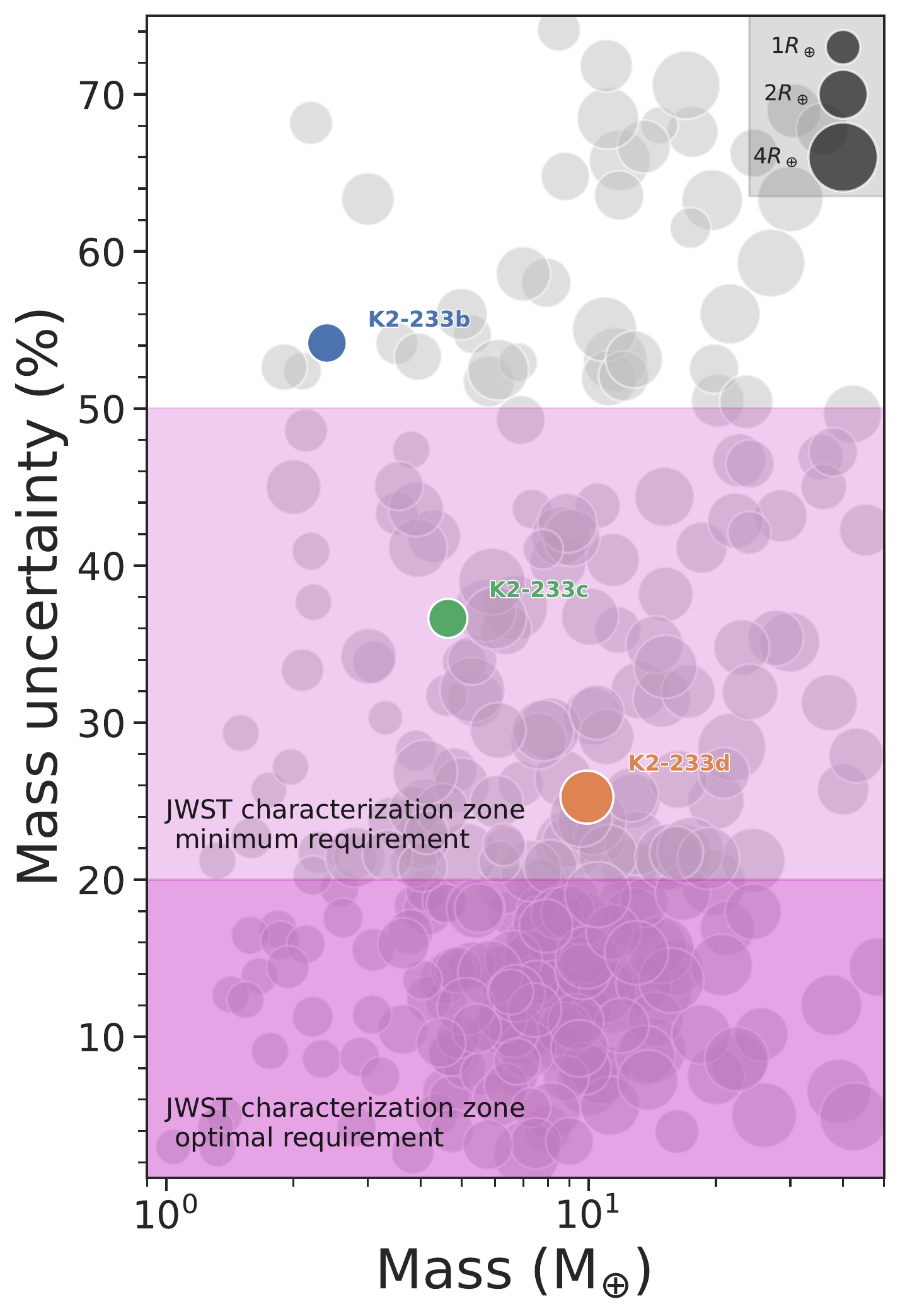}
    \caption{Mass uncertainty \emph{vs} mass plot for small exoplanets ($1 < R_{\rm p} < 4\, R_\oplus$ and $1 < M_{\rm p} < 50\, M_\oplus$). Grey circles correspond to planets with mass and radius measurements \citep[As in the NASA Exoplanet Archive on Feb 02, 2023;][]{NASAexoplanet}. 
    \targetbcd\ are shown with blue, green, and orange circles with corresponding labels identifying each planet. 
    Circles sizes correspond to planet radii, with the scale displayed in the upper right corner.
    The two shaded regions denote the minimum (lighter region) and optimal (darker region) regions for atmospheric characterisation with \emph{JWST}.
    }
    \label{fig:densitymass}
\end{figure}

Well determined masses and radii are crucial in order to perform further follow-up analysis of exoplanets, such as atmospheric characterisation. In particular,
\citet[][]{Batalha2019} show that a precision of at least $50\%$ in the mass determination is required to enable even a basic retrieval of atmospheric properties from transmission spectroscopic observations with \emph{JWST} or \emph{Hubble}, while a precision of $20\%$ or better is recommended to enable a more detailed atmospheric characterisation.
Figure~\ref{fig:densitymass} shows a mass \emph{vs} mass uncertainty plot for small planets. Thanks to the updated mass measurements obtained in this work, both \targetc\ and d now have sub-$50\%$ uncertainties on their masses, which meet the basic criterion of \citet[][]{Batalha2019}. We also computed the Transit Spectroscopy Metric \citep[TSM;][]{Kempton2018} of each planet, and report it in Table~\ref{tab:derived}. Due to their small radii, transmission spectroscopy observations of planets b and c would be challenging (for example, multiple transits of each planet would be required with JWST). Sub-Neptune \targetd\ is likely to host a H/He-dominated atmosphere. Based on this assumption, the TSM of planet d, while modest, is on par with that of K2-18b, which was recently characterised with WFC3/HST \citep{benneke2019}. We note that the recovered mild eccentricity of \targetd\ is not expected to be sufficient to induce observable time-variable atmospheric effects. Studies to date have shown that only high eccentricities have the potential to impact the atmospheric temperature profile and chemistry \citep[e.g.,][]{tsai2023,langton2008}.

Theory predicts a wide diversity for sub-Neptune atmospheres \citep{moses2013,guzman-mesa2022}. Observations to date have presented confounding results, some displaying clear water absorption features and others a featureless spectrum \citep{knutson2014,kreidberg2014,benneke2019}. Thick cloud and haze layers are thought to be more dominant for these smaller planets (than the Jupiter population), and this is the accepted explanation for dampened spectral features. At $T_\mathrm{eq, d}=$ \Teqd\ we would expect clouds to play a role in the atmosphere of \targetd, and this would have to be considered in the planning of atmospheric observations. That said, adding to the sample of observed sub-Neptunes is vital to understand this mysterious population, of which there is no analogue in the Solar system.
Furthermore, constraining the volatile content will inform formation process for close-in small planets \citep{bean2021}.

\subsection{On follow-up observations}

\target\ is a keystone target that can help to understand the evolution of multi-planetary systems, but additional follow-up observations are needed improve the precision of the planetary and orbital parameters. 
In particular, additional RV monitoring to refine the planet masses would be useful, as the results presented in this paper suggest that K2-233 b and c have significantly different masses, although their radii are nearly identical. This could arise from a discrepancy in composition between the two planets, which would challenge standard models of small-planet formation and evolution, but with the current data we cannot rule out that this is merely a statistical fluctuation.

How many more RV measurements would we need to improve the mass measurements of \target b and c to the point where we can confirm or rule out this tentative mass discrepancy? To answer this question, we used \texttt{citlalatonac} \citep[][]{pyaneti2} to simulate multiple realisations of additional HARPS-like observations and analysed them in the same way as we analysed the existing data in this paper.  
We used the GP hyper-parameters obtained in Table~\ref{tab:pars} to create the stellar signal as samples of a multi-GP. 
We simulated synthetic time-series of activity-induced RV, FWHM, and BIS and added the expected Keplerian signals for the three planets and sampled the data assuming \target\ is observed from La Silla with a maximum airmass of 1.5 in an intensive 3 months campaign, and added realistic white and instrumental red noise (based on the existing HARPS data). We then attempted to recover the stellar and planetary signals, using the same multi-GP framework and combining the existing and simulated data, as we have done with the available real data. By varying the number of synthetic observations, we established that we need around 50 epochs spread over 3 months to model the stellar signal robustly and to detect the signal of planet b to 3-sigma confidence or better, and improve the detection of planets c and d.

As well as spectroscopic data, more photometric data is also needed to refine the radii of the three planets and to keep track of the planet ephemerides. The latter is crucial in order to perform any transmission spectroscopy effort. 
Unfortunately, \target\ has not been observed by \tess\ and according to the Web \tess\ viewing tool\footnote{\url{https://heasarc.gsfc.nasa.gov/cgi-bin/tess/webtess/wtv.py}.} the star does not fall in any of the planned sectors until sector 69. This is due to the preferred orientation of \tess\ towards the poles and \target\ being a ecliptic target. We also checked if \target\ could be observable with the the CHaracterising ExOPlanet Satellite \citep[\cheops;][]{Benz2021}. 
According to the CHEOPS Observers Manual\footnote{\url{https://www.cosmos.esa.int/web/cheops-guest-observers-programme/cheops-observers-manual}.}, \target\ is observable each year for $\sim 70$\,d (between April and May) with an observing efficiency of 50\%\footnote{Each \cheops\ observation has a observing efficiency, which is the fraction of time on target  that is not interrupted due to its particular orbit.}. Thanks to its exquisite photometric precision, \cheops\ could be used to observe the transits of all \target\ planets, including the two small ones.
It is also tempting to think about ground-based photometric observations of \target. However, the shallow transits of \targetb\ and c are not observable from the ground. \targetd\ would be challenging but observable with state-of-the-art facilities, such as the Next Generation Transit Survey \citep[NGTS;][]{wheatley18ngts}.

\section{Conclusions}
\label{sec:conclusions}

We reanalysed the spectroscopic observations of \target\ published originally by \citet[][]{Lillo2020}. We used the activity indicators to constrain the stellar signal in the RV time-series using a multidimensional Gaussian Process approach. 
This led to an improvement of the detection of the Doppler signals with respect to the values reported in \citet[][]{Lillo2020}.
These improvements in the planetary mass measurements allow us to put better constraints on these planets' compositions. 

The outermost planet, K2-233 d, is consistent with a volatile-rich world, while K2-233 b and c are likely rocky, potentially hosting a volatile envelope of an insignificant mass fraction. Despite their similar sizes, K2-233 b and c seem to have different masses, suggesting a surprising discrepancy in composition between the two planets. However, further RV monitoring is needed to confirm this tentative discrepancy.
Having improved the precision of the planetary masses, we add the K2-233 planets to the small sample of young planets amenable to space-based transmission observations.

\section*{Acknowledgements}
{
This publication is part of a project that has received funding from the European Research Council (ERC) under the European Union’s Horizon 2020 research and innovation programme (Grant agreement No. 865624).
OB thanks Davide Gandolfi for his helpful insights on the discussion of the observability of \target\ with \cheops.
EG gratefully acknowledges support from the UK Science and Technology Facilities Council (STFC; project reference ST/W001047/1).
This work made use of \texttt{numpy} \citep[][]{numpy}, \texttt{matplotlib} \citep[][]{matplotlib}, and \texttt{pandas} \citep{pandas} libraries.
This work made use of Astropy:\footnote{\url{http://www.astropy.org}} a community-developed core Python package and an ecosystem of tools and resources for astronomy \citep{astropy1, astropy2,astropy3}. 
OB, AM, and BK would like to thank the Super Wrap place in Oxford that gave us the energy to write this paper at the best quality-price.
}

\section*{Data Availability}

The codes used in this manuscript are freely available at \url{https://github.com/oscaribv}.
The spectroscopic measurements that appear in Table~\ref{tab:harps} are available as supplementary material in the online version of this manuscript.

%
\bibliographystyle{mnras} 
\bibliography{refs} 
%

\begin{appendix}

\section{Characterising the stellar signal}
\label{sec:stellarsignal}

We perform a one-dimensional GP regression of different activity indicators to study the shape of the stellar signal in different time-series $\mathcal{A}_k$. We intend to analyse which time-series behave photometric- (they can be described only with $G$) and RV-like (they also need to be described with $\dot{G}$). 

We model the covariance between two times $t_i$ and $t_j$ for each time-series $\mathcal{A}_k$ with the function
\begin{equation}
    \gamma_{\rm 1D} = A^2 \gamma_{i,j},
\end{equation}
\noindent
where $A$ is an amplitude term, and $\gamma_{i,j}$ is the QP kernel given in eq.~\eqref{eq:gamma}.
We use the code \pyaneti\ \citep{pyaneti,pyaneti2} to create posterior distributions of the QP kernel hyperparameters.

We sample for 6 parameters in each run: four GP hyperparameters ($A$, \pgp, \lbe, \lbp), one offset, and a jitter term. 
We set wide uniform priors for all the parameters. In particular, for \lbe\ we set a uniform prior between 0 and 500 d, for \lbp\ between 0.1 and 10, and for \pgp\ between 8 and 12 d.
We sample the parameter space using the built-in MCMC sampler in \pyaneti. The posterior creation follows the same guidelines as in Sect.~\ref{sec:transitanalysis}. 

\begin{figure}
    \centering
    \includegraphics[width=0.45\textwidth]{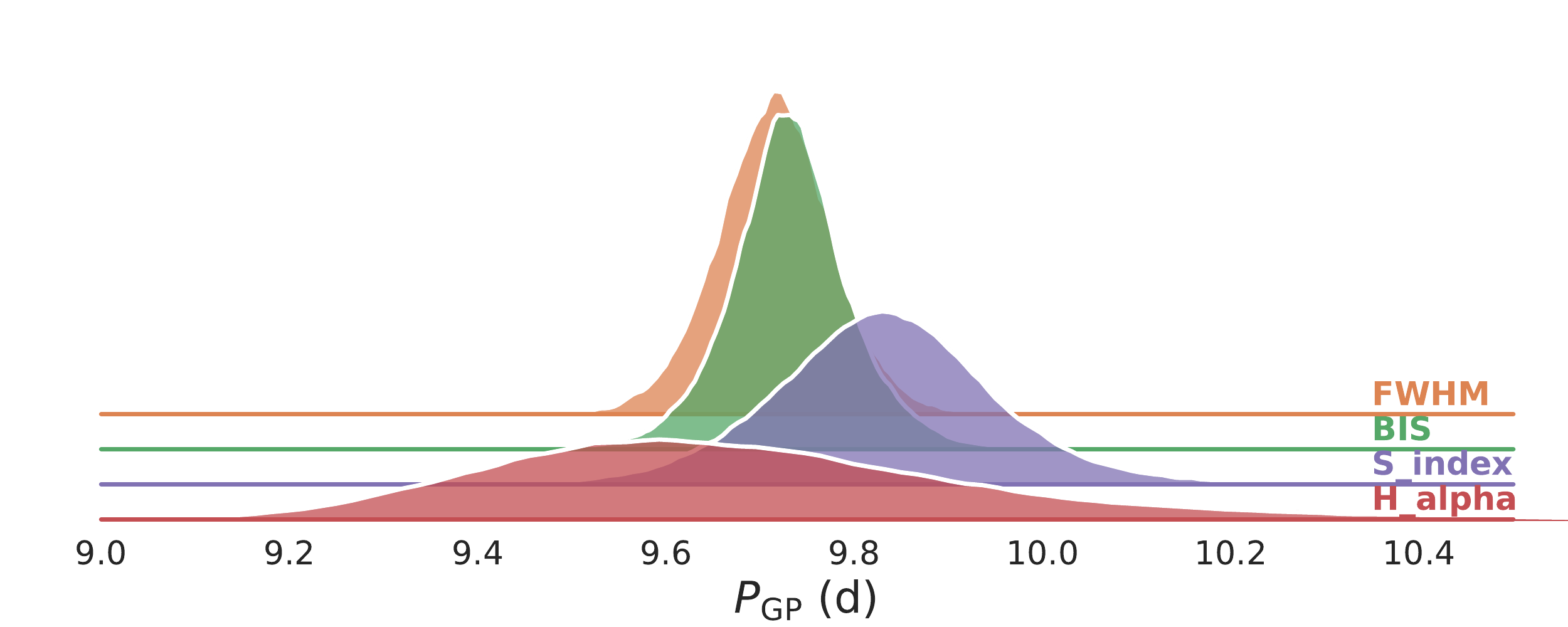}\\
    \includegraphics[width=0.45\textwidth]{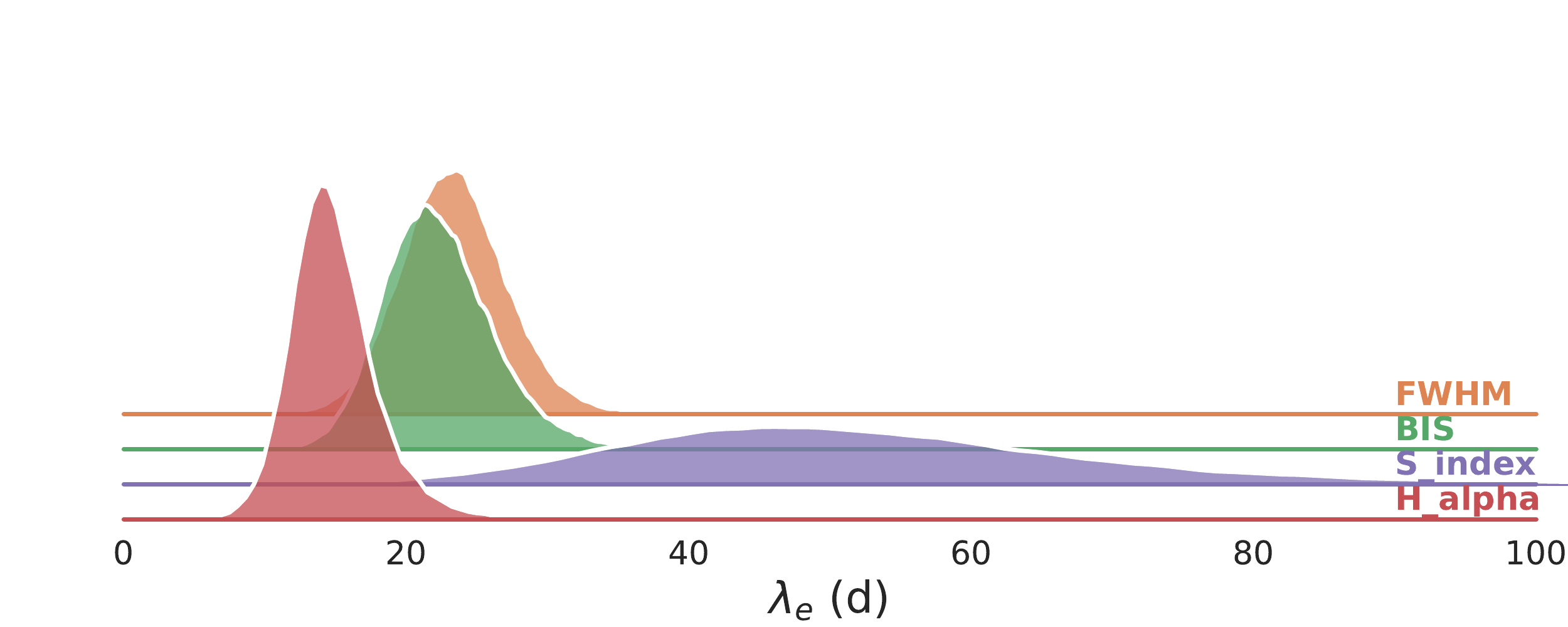}\\
    \includegraphics[width=0.45\textwidth]{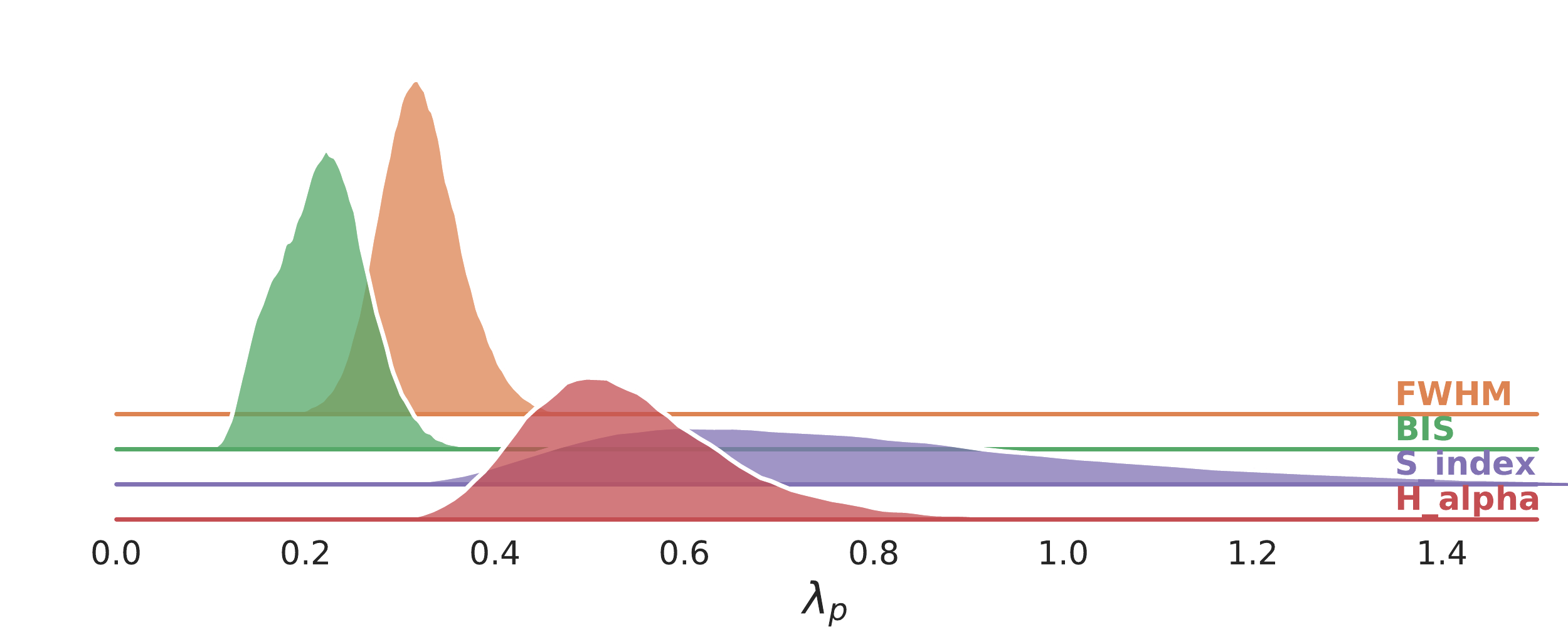}
    \caption{Posterior distributions for \pgp (top), \lbe\ (middle), and \lbp\ (bottom). Results for FWHM (orange), BIS (green), \sshk\ (purple), and \halpha\ (red) shown for each sub-panel.}
    \label{fig:posteriors1gp}
\end{figure}

\begin{figure*}
    \centering
    \includegraphics[width=\textwidth]{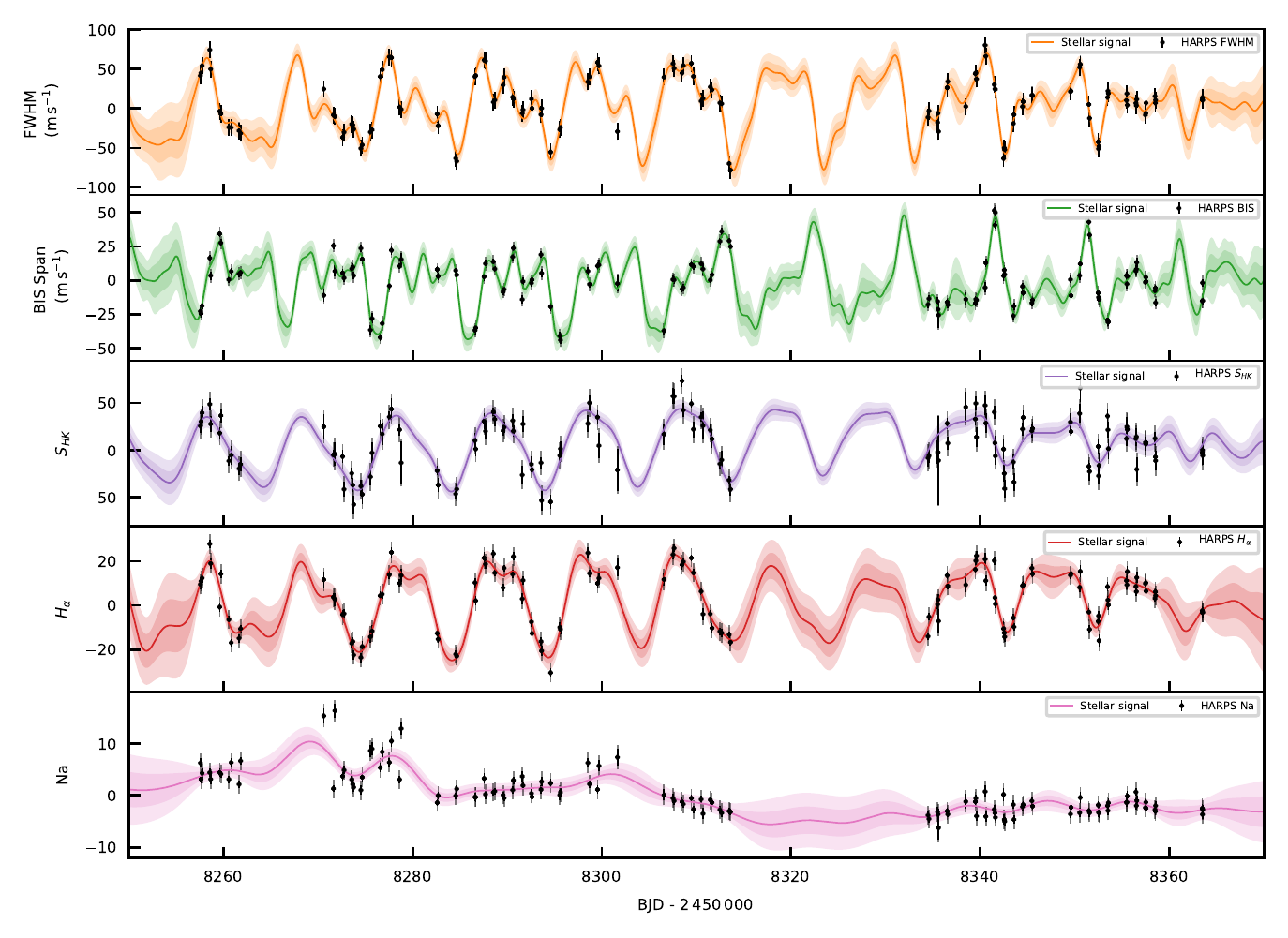}
    \caption{\target\ time-series for (from top to bottom) FWHM, BIS Span, \sshk, \halpha, and \sodium. 
    The corresponding measurements are shown with black circles with error bars with a semi-transparent error bar extension accounting for the inferred jitter. 
    Solid coloured lines show the corresponding inferred signal coming from our GP regression, while light coloured shaded areas show the one and two sigma credible intervals of the corresponding GP model.
    }
    \label{fig:timeseries1dgp}
\end{figure*}

Table~\ref{tab:1gphp} shows the inferred \pgp, \lbe, and \lbp\ hyperparameters for all time-series. Figure~\ref{fig:posteriors1gp} shows the inferred posterior distributions for the \pgp, \lbe, and \lbp\ parameters for all the modelled time-series, except for \sodium.
Figure~\ref{fig:timeseries1dgp} shows the data and inferred models for the different time-series.

\begin{table}
\begin{center}
\caption{Recovered hyperparameters for one-dimensional GP regressions. \label{tab:1gphp}} 
\begin{tabular}{lccccc}
\hline\hline
Time-series  & \pgp\ [d] & \lbe\ [d] & \lbp\   \\
\hline
FWHM & $9.72 \pm 0.06$ & $23.3_{-3.6}^{+3.8}$ & $0.32 \pm 0.04$ \\
BIS Span & $9.73 \pm 0.06 $ & $22.0_{-3.5}^{+4.1}$  & $0.22 \pm 0.05$  \\
\sshk & $9.83 \pm 0.11$ & $50.5_{-14.1}^{+18.4}$ & $0.75_{-0.22}^{+0.33}$ \\
\halpha & $9.65_{-0.22}^{+0.28}$ & $14.5_{-2.5}^{+3.1}$ & $0.53_{-0.09}^{+0.12}$ \\
Na & $9.54_{-0.81}^{+0.98}$ & $10.9_{-2.0}^{+2.1}$ & $2.23_{-0.70}^{+1.17}$ \\
\hline
\end{tabular}
\end{center}
\end{table}

The first thing to note is that the \sodium\ does not behave as a quasi-periodic signal in Figure~\ref{fig:timeseries1dgp}. 
\citet[][]{Diaz2007} described how the \sodium\ D1 and D2 lines could be used as chromospheric activity indicator for active late-type stars ( $B-V > 1.1$), typically M dwarfs \citep[][]{Gomes2011}, but not for earlier stars as \target\ ($B-V \approx 0.9$). This suggests that the \sodium\ is not a good activity indicator to constrain \target's stellar activity. 
This also shows how not all activity indicators are good tracers of the stellar signal that is also contained in the RV data.
From now on, we will exclude the \sodium\ time-series from our analysis. 

From Figure~\ref{fig:posteriors1gp} (and Table~\ref{tab:1gphp}) we can see that all time-series provide values for \pgp\ that agree well with each other.
These values are also in agreement with the stellar rotation period of $9.754 \pm 0.038$\,d reported by \citet[][]{David2018} from the \ktwo\ photometric analysis. 
This suggests that all these activity indicators are good tracers of the stellar rotation period. 

When we look at the recovered values for the inverse of the harmonic complexity, \lbp, we can see that we have different values for different activity indicators (see Figure~\ref{fig:posteriors1gp}). 
First we note that the recovered \lbp\ for the BIS time-series is consistently smaller (i.e., higher harmonic complexity) than the other activity indicators. 
As discussed in \citet[][]{Aigrain2012} and  \citet[][]{pyaneti2,Barragan2022}, it is expected that the BIS time-series is \emph{RV-like}, meaning that it depends on the change of location of the active regions from the red- to the blue-shifted stellar hemisphere, and vice-versa.
We then see that the FWHM peaks around a value of 0.3, implying a lower harmonic complexity than the BIS time-series (see Table~\ref{tab:1gphp}). This behaviour is also expected, given that previous analyses suggest that these time-series are good tracers of the areas covered by active regions on the stellar surface \citep[e.g.,][]{Oshagh2017}. For this reason, we call these kinds of activity indicators \emph{photometry-like}.
Is it worth noting that these photometric and RV-like behaviours are visible in Fig~\ref{fig:timeseries1dgp}. The BIS time-series behaves visually as the derivative of the FWHM time-series. This is expected and it is the base of $FF'$ models as the one used in the multi-GP approach \citep[see][]{Aigrain2012,Rajpaul2015,pyaneti2}.
Finally we note that the recovered process for the chromospheric activity indicators, \halpha\, and \sshk\, have a relatively low harmonic complexity (i.e. values of \lbp\ that are relatively large). This is unexpected given that both, \halpha\, and \sshk\, are expected to be photometry-like activity indicators.
Therefore, we would have expected that they would behave similar to FWHM. 
Moreover, from Figure~\ref{fig:timeseries1dgp} we can see that data in these two time-series have a relatively high white noise in comparison with the amplitude of the recovered signal. This significant white noise can be explained because the star is relatively faint ($V \sim 10.88$ mag).
We speculate that the relatively high white noise in these time-series degrades our ability to recover complex patterns (i.e., higher harmonic complexity) of the stellar signal. This means that even if the underlying process has a significant harmonic complexity, we cannot recover it due to the sub-optimal data.

We then analyse the long-term evolution time-scale, \lbe\ (see Table~\ref{tab:1gphp}). 
Both, the FWHM and the BIS time-series give a \lbe\ that peaks around $\sim 22$\,d, that is twice the \pgp. This implies that the local coherence of the periodic signal last two rotation periods. When analysing the recovered \lbe\ for the chromospheric activity indicators, we notice that they also produce different values when compared with the CCF ones (see Table~\ref{tab:1gphp}). 

A priori, we expect that FWHM, \sshk, and \halpha behave as photometric like activity indicators and that BIS behaves as RV-like activity indicators. So we would have expected that all activity indicators trace similar \pgp\ and \lbe; while we would suppose a higher harmonic complexity for RV-like activity indicators. This is the general tendency that we observe, but with significant annotations. First, it seems that all activity indicators are good tracers of the rotational period of the star. This is expected given that the stellar rotation is the dominant imprint of the stellar signal in the spectroscopic time-series. 
Second, more subtle characteristics of the stellar signal, that can be traced with \lbe\ and \lbp\, seem to vary between different time-series. 
We hypothesise that this difference is caused by a degradation of the recovered stellar signal due to difference in data quality for each time-series.
But we note that this difference can also be due to the fact that activity indicators are mapping different areas of the stellar surface, as well as we reaching the limits of the model based on a $FF'$-like relation between the RVs and activity indicators. 

From these analysis and discussion we will assume that, for this specific system, the CCF diagnostics are better activity indicators than the chromospheric data to characterise the stellar signal in the spectroscopic time-series.
We therefore use the FWHM and BIS time-series for the multi-GP analyses presented in Sect.~\ref{sec:multigps}.



\end{appendix}

\bsp	
\label{lastpage}
\end{document}